\documentclass[superscriptaddress, amsmath,amssymb, aps,prb,floatfix,twocolumn,longbibliography]{revtex4-1}
\usepackage{graphicx}
\usepackage{dcolumn}
\usepackage{bm}
\usepackage[utf8]{inputenc}
\usepackage[T1]{fontenc}
\usepackage{mathptmx}
\usepackage{enumitem}
\usepackage{caption}
\usepackage{subcaption}
\usepackage{gensymb}
\usepackage[version=4]{mhchem} 
\usepackage[english]{babel}
\usepackage{float}

\usepackage{placeins}

\usepackage{microtype}

\begin{document}

\preprint{AIP/123-QED}

\title[SAXS magnetic vapor annealing setup]{Solvothermal vapor annealing and environmental control setup with adjustable magnetic field module for GISAXS studies.
}

\author{Christian Kjeldbjerg}
\affiliation{IMFUFA, Department of Science and Environment, Roskilde University, DK-4000 Roskilde, Denmark}
 
\author{Bo Jakobsen}
\affiliation{IMFUFA, Department of Science and Environment, Roskilde University, DK-4000 Roskilde, Denmark}

\author{Miriam Varón}
\affiliation{Department of Physics, Technical University of Denmark, DK-2800 Lyngby, Denmark}

\author{Kim Lefmann}
\affiliation{Nanoscience Center, Niels Bohr Institute, University of Copenhagen, DK-2100 Copenhagen \O , Denmark}

\author{Cathrine Frandsen}
\affiliation{Department of Physics, Technical University of Denmark, DK-2800 Lyngby, Denmark}

\author{Dorthe Posselt$^*$}
\affiliation{FRUSTMI, IMFUFA, Department of Science and Environment, Roskilde University, DK-4000 Roskilde, Denmark}

\date{\today}

\begin{abstract}
A compact, modular environmental control and solvothermal vapor annealing chamber designed for maintaining a controlled atmosphere with regard to solvent humidity and temperature is presented. The setup allows \textit{ex situ} and \textit{in situ} grazing incidence small-angle X-ray scattering (GISAXS) investigations of thin film self-assembly and reorganization. Its modular slotting system enables stable reconfiguration, including the integration of an adjustable magnetic field module. The temperature is maintained via a water-based heating and cooling loop supplemented by resistive elements, and the solvent vapor environment is regulated using a commercial controlled mixing and evaporation unit. The performance of the setup is validated through measurements of fill and quench times together with magnetic field mapping with Gauss meter measurements and finite element simulations. Further, the versatility of the setup is demonstrated with four research examples using the chamber for solvothermal vapor annealing of block copolymer thin films together with lab-based \textit{ex situ} and \textit{in situ} GISAXS measurements. The portable new design offers robust environmental control and flexibility for advanced thin film investigations both in the lab and at large scale facilities. The design can be adapted for grazing incidence small-angle neutron scattering, GISANS.

\end{abstract}

\maketitle

\section{\label{sec:intro}Introduction}
A well-defined sample environment with high precision control of relevant parameters during small-angle X-ray scattering in transmission geometry (SAXS) and in grazing incidence geometry (GISAXS) often requires control of the atmosphere in the sample chamber (e.g., vacuum, inert gas, a certain level of water or solvent humidity), and the sample temperature. Typical soft matter thin film samples, which are mounted in direct contact with the sample environment atmosphere, include for instance block copolymers (BCPs), liquid crystals, and stacked phospholipid bilayers. Liquid crystals, which exhibit temperature-dependent phase behavior, require high precision temperature regulation to capture transitions between mesophases, e.g., from nematic to smectic phases\cite{Singh2000}. Phospholipids, as found in biological membranes, undergo phase changes depending on the temperature and hydration level\cite{Nagle2000}. BCPs require thermal control to study their microphase separation\cite{Bates1990}. In addition to thermal and humidity control, magnetic BCP thin films and hybrid BCP/magnetic‑nanoparticle films can be aligned by external magnetic fields integrated into the sample environment, resulting in ordered nanostructures such as cylinders\cite{Gopinadhan2014} and nanowires\cite{Fang2007, Yao2014}. 
Thin BCP films prepared by, e.g., spin coating typically have a number of kinetically trapped defects, hence calling for annealing of the structure. For many BCP samples, thermal annealing is slow and potentially harsh and does not offer a large degree of control of the ordering process. Solvent vapor annealing (SVA), on the other hand, has emerged as a versatile technique where setups of increasing sophistication beyond simple jar annealing (i.e., a closed container with a liquid reservoir) offer a large degree of control over the annealing process\cite{Knoll2004, Sinturel2013, Posselt2017}. Solvothermal vapor annealing (STVA), where the temperature of the SVA setup can be raised somewhat above room temperature, facilitates the formation of highly ordered BCP thin film structures both by reducing the glass transition temperature (T$_g$) of the BCP constituent polymers due to uptake of solvent and by enhancing the molecular mobility of the BCP. 
Various characterization and environmental control components are often added to SVA setups such as devices for spectral reflectometry \cite{Di2009,Gotrik2012,Emerson2013,Gotrik2013,Gu2013,Sepe2014,Kim2014,Sepe2016,Nelson2018,Evans2018,Hulkkonen2019,Ma2023}, ellipsometry\cite{StenbockFermor2014,Cheng2019,Efremov2018,Bilchak2020}, and cooling/heating elements for STVA.\cite{Kim2013,Gotrik2013,Dinachali2015,Cummins2016,Kim2016,Berezkin2018,Efremov2018,Cheng2019}.
Setups have also been developed for \textit{in situ} GISAXS experiments at synchrotron facilities\cite{Di2009,Gu2013,Sepe2014,Gowd2014,Chavis2015,Berezkin2018,Dolan2018} making real-time tracking of structural changes in the films possible.

In this paper, we present an environmental control and STVA setup designed for annealing experiments including \textit{in situ} GISAXS studies at both synchrotron and laboratory sources. The setup has a flexible modular design enabling a high degree of control of the environmental conditions to which the sample is exposed, typically composition of sample chamber atmosphere, temperature and magnetic field. The setup presented in this paper is a complete redesign of a previous setup from our group \cite{Ariaee2023} and has been developed to meet new requirements, including a higher degree of environmental control, a faster response time with regard to atmospheric changes and the possibility of applying a magnetic field. The new design is more stable and flexible with an increased ease of accessibility, for instance when changing the sample. The chamber design can be adapted for grazing incidence small-angle neutron scattering (GISANS) measurements, broadening its applicability to include neutron-based structural and magnetic investigations.

\section{Solvent vapor annealing and environmental chamber design}\label{sec:ExpSetup}

\subsection{Basic principles}
The STVA setup presented here is constructed mainly with the purpose of studying BCP thin films, the main experimental techniques being atomic force microscopy (AFM) before and after annealing in the STVA chamber and GISAXS with the sample placed in the STVA chamber and annealing performed either before transfer of the STVA chamber (or the annealed sample alone) into the X-ray beam (\textit{ex situ} GISAXS) or during annealing (\textit{in situ} GISAXS). 

Bulk BCP phase behavior is determined by the chemistry of the different polymers that make up the individual BCP blocks and are parametrized by the Flory-Huggins interaction parameter $\chi$, and topological parameters. These are the relative size of the different blocks bonded together through covalent bonds; the overall size of the BCP; and the overall BCP architecture (e.g., linear or star BCPs\cite{Bates1990, Bates2016}). For low values of $\chi$ (or correspondingly high temperatures, $\chi \propto T^{-1}$), the BCP forms a disordered melt. Lowering the temperature results in a disorder-order transition, the so-called microphase separation, where the different blocks segregate as much as bonding constraints allow into separate microdomains forming an overall ordered structure such as lamellae; cylinders on a hexagonal lattice; spheres on a cubic lattice\cite{Bates1990}; or more complex structures, e.g., gyroid or Frank-Kasper phases\cite{Hajduk1994, Lee2010}. For thin BCP films, additional parameters come into play, such as interaction with the substrate and the film thickness (commensurate or non-commensurate with the characteristic BCP microdomain size)\cite{Hamley2009}. As detailed elsewhere\cite{Sinturel2013, Posselt2017}, the SVA process relies on the uptake of solvent to screen polymer–polymer interactions leading to an effective reduction in $\chi$ together with a reduction in the glass transition temperature, $T_g$, due to the small solvent molecules effectively acting as plasticizers. When the thin film is exposed to solvent vapor, uptake of solvent results in swelling of the film. The swelling process is typically monitored through the swelling ratio, $SR(t)$, the ratio of the swollen film thickness, $d(t)$, at a given time $t$ to the initial dry film thickness, $d_0$:
\begin{equation}
    SR(t) = \frac{d(t)}{d_0},
\end{equation}
 Reorganization of the BCPs towards the thermodynamic equilibrium structure (in the swollen state) is a complex process characterized by several time- and length scales. When the desired structure and degree of order is reached, the thin film is dried. It can either be quenched by fast drying or allowed to slowly relax toward the dry equilibrium state. The details of the full process can be controlled and fine-tuned experimentally by choice of solvent (selective or non-selective for the different polymer blocks), solvent vapor pressure during annealing, swelling protocol, drying rate and annealing temperature. Sophisticated protocols involve the use of several solvents\cite{Jung2020}. In the following, we use 'relative humidity' or 'solvent humidity' to express the degree of wetness in the chamber atmosphere with respect to the relevant solvent vapor, i.e., the concentration of solvent gas in the total chamber gas with 100$\% $ referring to a chamber atmosphere saturated with the relevant solvent gas. 

GISAXS allows for the probing of thin film structures along the film normal and laterally in the plane perpendicular to the film normal. This is achieved by the incident X-ray beam illuminating the sample at a grazing angle, $\alpha_{i} \sim0.1\degree$. This way, a large footprint ensures a large illuminated thin film volume. The critical angle, $\alpha_{c}$, of a given film material defines the angle above which the X-ray beam will begin to penetrate the material. For $\alpha_{i}$ < $\alpha_{c}$ the full beam is reflected and is marked by an intense spot on the 2D detector behind the sample, positioned at an X-ray beam exit angle equal to $\alpha_{i}$. A schematic of GISAXS scattering geometry can be found in Supplementary Material Fig.~S1. A substrate is typically chosen to have a higher critical angle, $\alpha_{cs}$, than the film material under study, $\alpha_{cs}>\alpha_c$, and three regimes are identified\cite{MllerBuschbaum2009, Posselt2017, Smilgies2022}:
\begin{itemize}
    \item{\makebox[2.5cm]{$\alpha_i<\alpha_c$:\hfill} Evanescent regime}
    \item{\makebox[2.5cm]{$\alpha_c<\alpha_i<\alpha_{cs}$:\hfill} Dynamic regime}
    \item{\makebox[2.5cm]{$\alpha_i>\alpha_{cs}$:\hfill} Kinematic regime}
\end{itemize}
In the evanescent regime, only the outermost few nm of the film surface is probed by the evanescent wave penetrating into the film. In the dynamic regime, the beam illuminates the whole film interior, and correspondingly, reflection and refraction of the beam at interfaces must be taken into account, resulting in complex dynamical effects. The scattering parallel to the film surface, i.e., along $q_y$, (see Fig.~\ref{fig:PSPB60GISAXS} for an example) gives information on lateral film structuring and is interpreted using standard SAXS scattering theory, i.e., form and structure factors~\cite{Jeffries2021}. Scattering in the scattering plane, defined by the incoming and the specularly reflected beam i.e., along $q_z$, gives information on the electron density depth profile in the film. The quantitative interpretation of this part of the scattering is complex, requiring the use of the distorted Wave Born approximation\cite{BuschP.2006GsXs}. In the kinematic regime, when $\alpha_i>\alpha_{cs}$, the intensity of the reflected signal drops sharply and modeling of the scattered intensity is typically much simplified~\cite{Lazzari2002, MllerBuschbaum2009}. 

Examples of the use of GISAXS to characterize restructuring processes in thin films using the STVA setup either \textit{ex situ} or \textit{in situ} are given at the end of the paper. First, we list the design criteria and describe the new design of the STVA setup while emphasizing the differences compared to our previous setup\cite{Ariaee2023}. The setup is then characterized with respect to filling the STVA chamber with solvent gas (swelling) and inert gas (drying), respectively, and the magnetic field available during STVA is characterized.

\subsection{STVA chamber design criteria}

The STVA setup consists of four modules: 1) the STVA chamber with accompanying exchangeable modular drawers, 2) a system for solvent vapor and nitrogen gas delivery, 3) a UV cell for measurement of solvent vapor concentration in the chamber exhaust gas (SVC), and 4) a commercial UV-VIS-NIR light spectral reflectometer for \textit{in situ} measurements of the thin film thickness. 

The STVA chamber is designed to meet the following criteria:

\begin{enumerate}
    \item \textbf{GISAXS compatibility:} The chamber must be suitable for both laboratory-based and synchrotron GISAXS. It must be portable and lightweight and possible to mount on a goniometer allowing control of $\alpha_i$ with high precision ($\pm 0.002\degree$), tilt of the sample around the X-ray beam axis and possibly also rotation around the sample normal, both with high precision. 
    \item \textbf{Real-time thickness tracking:} It should be possible to connect the optical fiber of an UV-VIS-NIR spectral reflectometer to a chamber window in a stable and reproducible way, thus separating the fiber-optics from a potentially harsh atmosphere in the sample chamber and allowing reliable \textit{in situ} film thickness tracking.
    \item \textbf{Small sample chamber:} The chamber should have a compact design with a small free volume to allow fast environmental changes during solvent treatment and drying of thin films. 
    \item \textbf{Sample table and gas inlet:} The chamber should be equipped with a sample table that allows stable sample mount and gas inlet in a well-defined way (in time and space), with respect to all parts of the sample surface.
    \item \textbf{Temperature control:} It must be possible to heat and cool the sample and maintain high-precision temperature control during the \textit{in situ} experiments.
    \item \textbf{Magnetic field compatibility:} The chamber should allow the sample to be mounted in a magnetic field with well-defined direction relative to the sample normal thus enabling studies involving magnetically sensitive materials.
\end{enumerate}

\begin{figure}[!h]
     \centering
     \begin{subfigure}[b]{0.23\textwidth}
         \centering
         \includegraphics[width=\textwidth, trim= 40 40 40 40,clip]{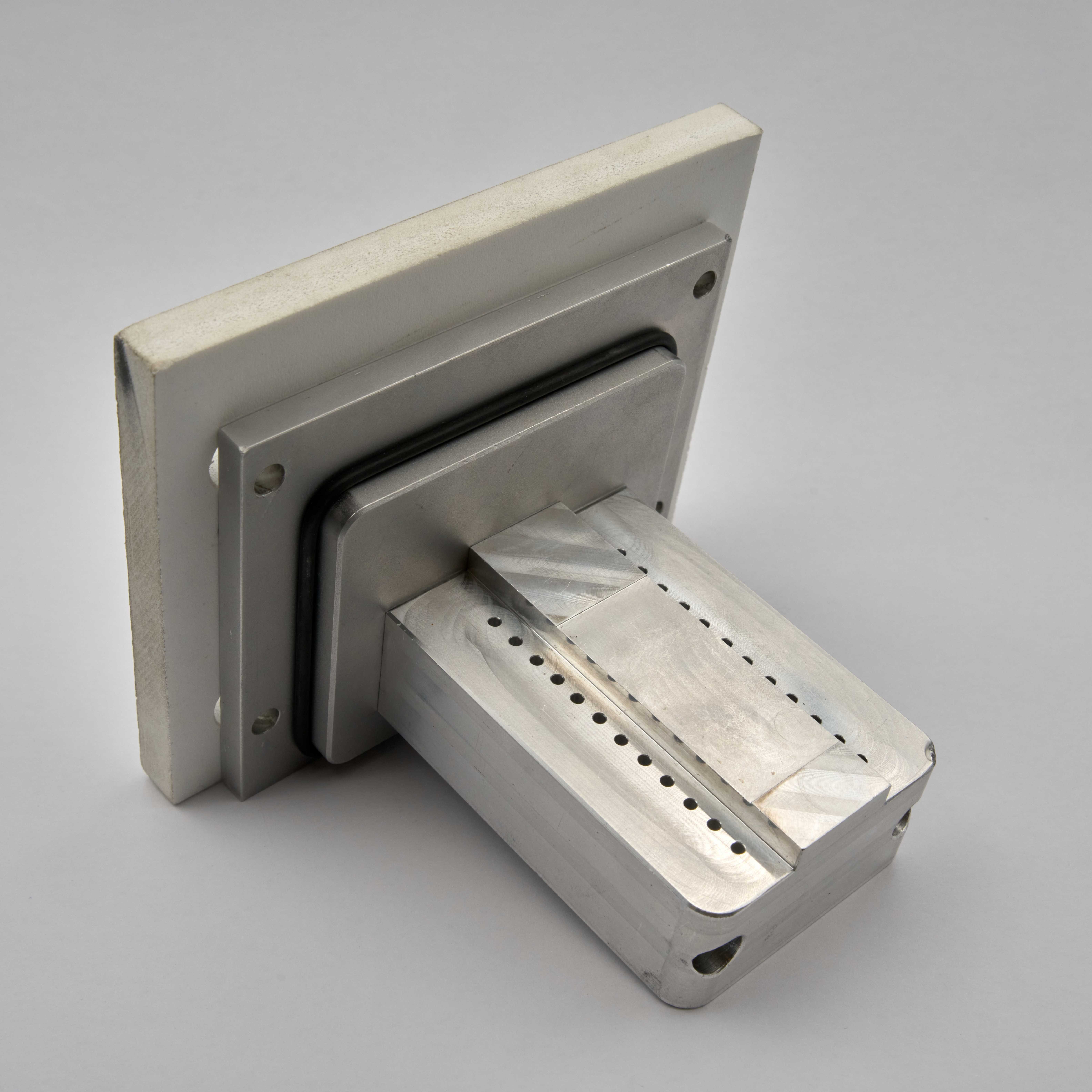}
         \caption{}
         \label{fig:shelf_base}
     \end{subfigure}
     \hfill
     \begin{subfigure}[b]{0.23\textwidth}
         \centering
        \includegraphics[width=\textwidth, trim= 40 40 40 40,clip]{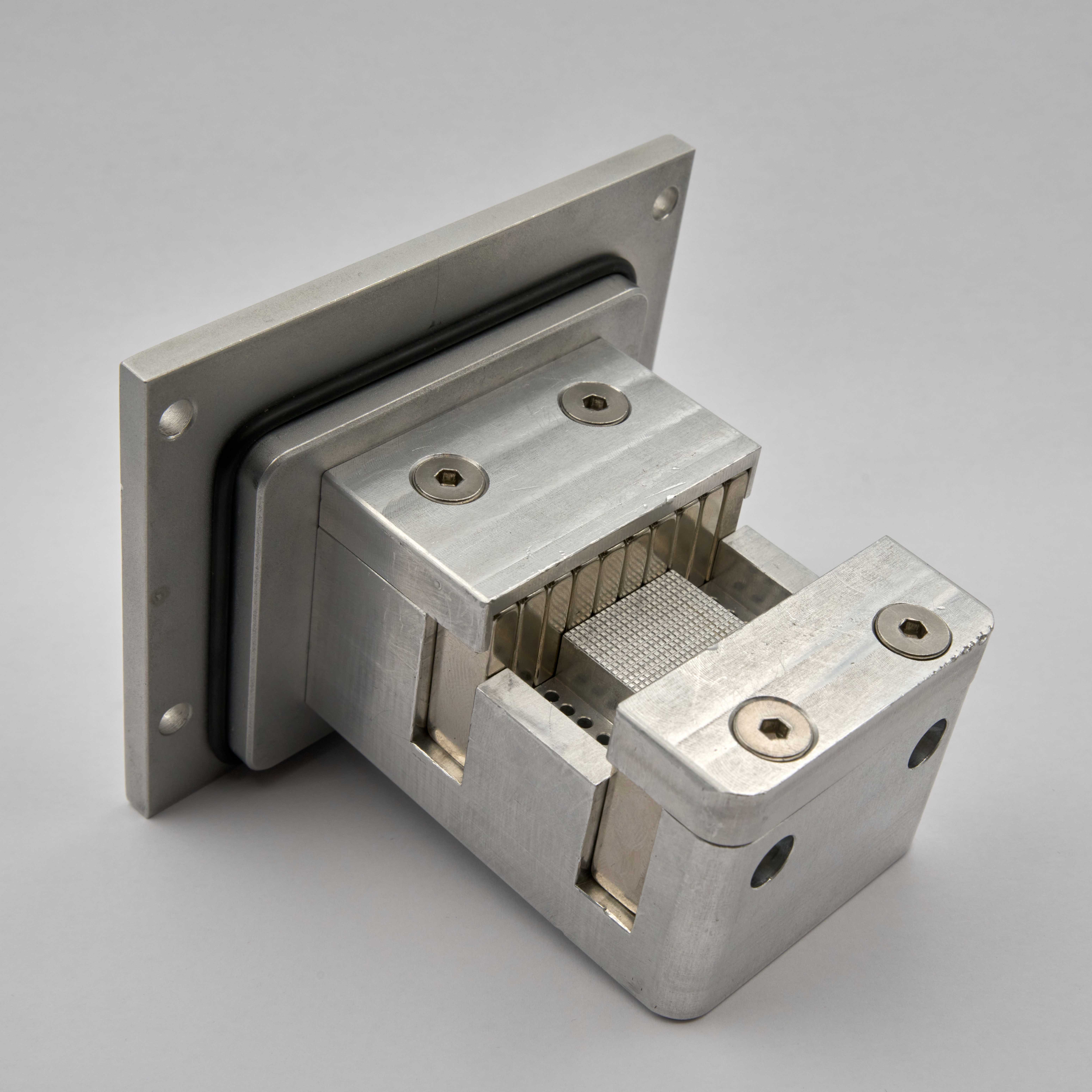}
         \caption{}
         \label{fig:shelf_magn}
     \end{subfigure}
     \begin{subfigure}[b]{0.23\textwidth}
         \centering
         \includegraphics[width=\textwidth, trim= 40 40 40 40,clip]{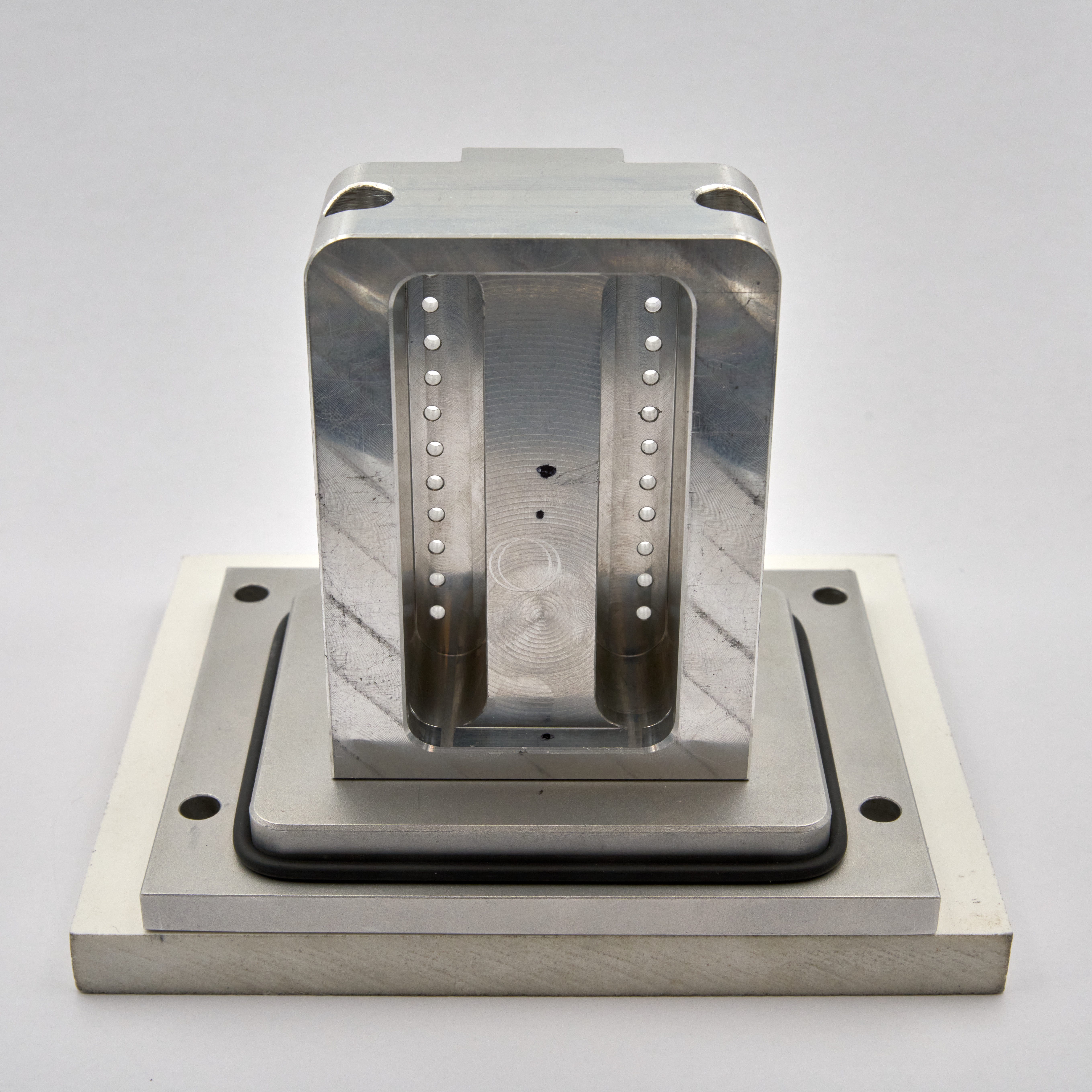}
         \caption{}
         \label{fig:mixing}
     \end{subfigure}
     \hfill
     \begin{subfigure}[b]{0.23\textwidth}
         \centering
        \includegraphics[width=\textwidth, trim= 40 40 40 40,clip]{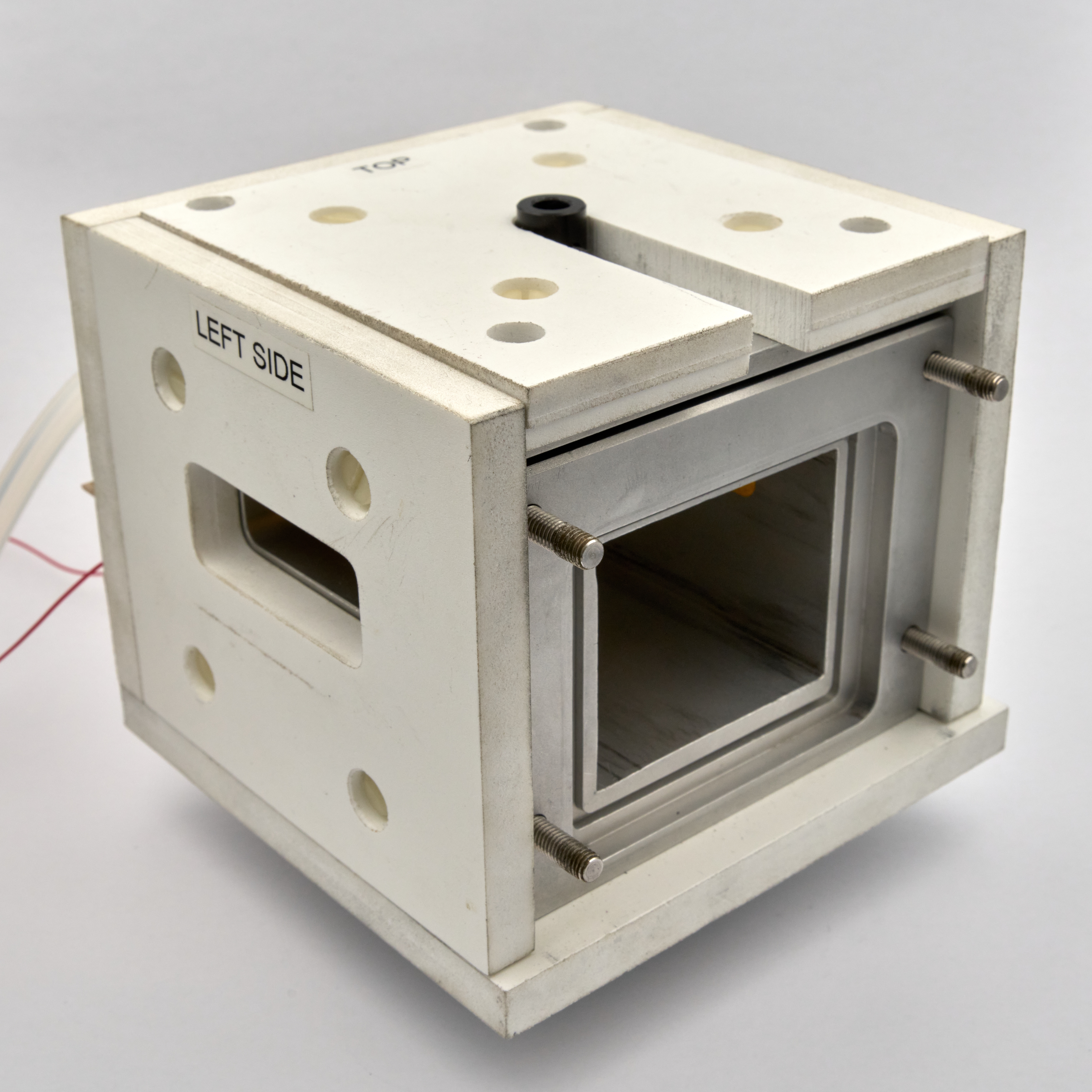}
         \caption{}
         \label{fig:opening}
     \end{subfigure}
        \caption{(a) Base drawer with large sample stage, (b) Magnetic drawer featuring a smaller sample stage designed for integrating permanent magnets, seen installed here on either side of the sample stage, to produce adjustable in‐plane and out‐of‐plane fields, c) Underside of base drawer featuring a mixing cavity for the incoming vapor  and d) Chamber housing covered in insulation sheets and displaying the side with the opening where drawers are slotted in. An X-ray transparent window made of Kapton is also seen on the figure.}
        \label{fig:shelves}
\end{figure}

The STVA chamber is constructed with stability and ease of access in mind. This has led to a drawer design that allows easy sample change by inserting a drawer with the sample table into a slot on the side of the chamber - see Fig.~\ref{fig:shelves}. This modular approach enables the option of changing the sample environment by changing the design of the drawer that goes into the slot. A Viton O-ring sits between the drawer backplate and the chamber house and is held in place with four brass thumb screws, ensuring a tight seal. The drawers and chamber housing are CNC-machined from aluminum, making them solvent resistant and non-magnetic to not influence the magnetic fields generated by permanent magnets installed in one of the drawers available for experiments. The chamber is insulated on all six sides using Vekaplan-S rigid PVC foam insulation sheets held in place by plastic screws. Fig.~\ref{fig:shelves} shows photos of the base drawer (\ref{fig:shelf_base}), the magnetic drawer (\ref{fig:shelf_magn}), the underside of the base drawer (\ref{fig:mixing}) together with the chamber housing with insulation sheets (\ref{fig:opening}). The base drawer has a sample table of $75 \times 20$~mm$^2$ that allows STVA of multiple samples simultaneously, while the magnetic drawer has a sample table of $22 \times 22$~mm$^2$. Both drawers have a mixing cavity on the backside, as shown in Fig.~\ref{fig:mixing}, where gas is let in before entering the chamber proper through a number of holes evenly distributed on either side of the sample stage. This design promotes a uniform distribution of the gas around the sample and ensures that all parts of the sample are exposed to vapor at the same time without any delays between the different sides of the sample (as observed in very early versions of STVA setups with gas inlet on one side). The magnetic drawer has room for an array of NdFeB permanent bar magnets on either side, each of which can be replaced by an aluminum dummy. By changing the orientation of the magnets both in-plane and out-of-plane field directions with respect to the sample surface can be produced. The compact design of the cell makes exchange of chamber gas faster than in our previous STVA setup\cite{Ariaee2023} - see Tab.~\ref{table:fill}. The current version has a free volume of $\sim$92~cm$^3$ with the base drawer and $\sim$45~cm$^3$ with the magnetic drawer while the older chamber had a volume of $\sim$375~cm$^3$. Kapton windows ($\sim$50~$\mu$m), on each side of the chamber allows for \textit{in situ} X-ray measurements. Supplementing (e.g., positioned below the Kapton windows), or substituting the X-ray transparent windows with sub-mm thick aluminum windows will allow the same STVA setup to be used for GISANS experiments. It may be advantageous to use a sample table which allows larger sample areas for neutron scattering than typically used for the X-ray scattering version. The chamber can also be used for transmission SAXS with the sample mounted upright on the sample table using a sandwich holder with a suitable large base allowing stable upright positioning. 

A water cooling / heating loop is drilled inside the base of the chamber and is connected to a Julabo DYNEO DD-200F circulator via insulated tubing. Two resistive heating elements (Polyimide Thermo-foil Heater, 65~W, Minco Co.) located on the sides of the chamber allows for enhanced heating capability. A temperature probe (PT100) is glued directly to the chamber wall for temperature measurement and regulation. The chamber is designed for use with the surrounding atmosphere at ambient pressure.

\subsection{Controlled evaporation and mixing system}
The solvent gas is produced using a commercial controlled evaporation and mixing (CEM) system from Bronkhorst High-Tech B.V. The CEM consists of an evaporator (CEM Evaporator W-102 A) together with a mini CORI-FLOW 20~g/h Coriolis liquid flow meter to generate solvent vapor at specified humidity, temperature and flow rate. The overall setup is composed of 3 main components:

\begin{enumerate}
    \item A liquid solvent reservoir is held in a 250 ml Borosilicate glass container 
 (Duran pressure plus+ to safely maintain pressures up to $\sim$1.5 bars) connected to a nitrogen gas inlet which supplies $\sim$0.7~bar overpressure. Through pressurization of the reservoir, a pure liquid solvent stream is extracted and transferred to the CEM mixing valve.
    \item The Coriolis liquid flow meter (20~g/h mini CORI-FLOW Bronkhorst High-Tech B.V.) measures and controls the liquid flow into the CEM mixing valve. The mixing valve controls the injection of a small amount of liquid into the steady carrier nitrogen gas flow of up to 200~sccm (cm$^3$ per minute at standard temperature and pressure), supplied by a mass flow controller from Bronkhorst (MFC). The resulting aerosol is then transferred to the evaporator.
    \item The liquid-gas aerosol is evaporated in the temperature-controlled evaporation chamber of the CEM. The resulting gas mixture then passes through a tube into the STVA chamber. 
\end{enumerate}    
The liquid flow through the flow meter is regulated together with the temperature of the evaporation chamber to obtain the desired solvent concentration (humidity), gas flow into the STVA chamber and final temperature of the gas mixture. The required set point liquid flow is calculated using an online software (FLUIDAT, from Bronkhorst). Henceforth, all relative humidities quoted for nitrogen–solvent mixtures correspond to the values calculated using FLUIDAT. By setting the temperature in the evaporation chamber slightly higher than the target temperature, condensation in the connecting tubes is avoided. A second MFC is connected to the STVA chamber via a separate inlet and can supply up to 400~sccm nitrogen flow during drying.

\subsection{Solvent vapor concentration unit and spectral reflectometer}
 The working principle of the SVC unit for tracking solvent concentration in the exhaust gas from the STVA chamber is as follows: The exhaust gas flows through a cell, where a mercury lamp providing 254 nm UV radiation illuminates a photodetector placed at the opposite end of the cell. The resulting absorbance of the exhaust gas, $A$, is calculated as:
\begin{equation}
    A = -\log_{10}T = -\log_{10}\frac{I}{I_0},
\end{equation}
where the transmittance, $T = I/I_0$, is the ratio between the UV intensity, $I$, transmitted through the solvent containing exhaust gas, and the intensity, $I_0$, transmitted through pure nitrogen exhaust gas. The SVC unit works with solvents absorbing 254 nm UV light, typically solvent molecules with $\pi$-bonding, e.g., toluene or acetone.

Building on our previous design\cite{Ariaee2023}, the SVC unit has now been upgraded to include temperature control. A heating wire heats the SVC cell to avoid condensation of solvent inside the cell during SVA. The SVC unit is operated by remote control from a computer outside the experimental area via a microcontroller (Arduino SA) connecting to a PT100 temperature probe attached to the cell housing.

We use a UV-VIS-NIR commercial spectral reflectometer (NanoCalc-XR, Ocean Optics Ltd.; repeatability 0.3 nm) to continuously monitor film thickness during swelling and drying. The fiber-optic probe is mounted on top of the STVA chamber, directing light through a sapphire window onto the sample table surface at the standard position of the thin film sample. In the current setup, samples are loaded through the side of the STVA chamber, in contrast to the former design, where the loading took place through the top part where the fiber-optics are mounted. With the new design we have improved the stability and the reproducibility of the thickness measurements. The sample-probe distance can be adjusted to optimize the reflected intensity enabling the accommodation of various substrate thicknesses and sample stage heights. Calibration is performed by providing the spectrum from a reference sample, typically a bare Si wafer, and a dark measurement i.e.~a measurement of any stray reflections originating in the light optical path. The dark measurement is performed by covering the sample stage with a piece of IR flock sheet (Musou Black USA, Costa Mesa, CA, USA), which has a reported 0.3\% average reflectance in the visible spectrum thus largely ensuring that only internal reflections are sampled during calibration. Finally, a polystyrene thin film sample with known thickness is used as a standard sample for checking the thickness measurement setup. Analysis of the measured reflectance requires a model for the film measured in terms of the number of layers in the film with thickness and refractive index for each layer\cite{MacLeod2010}.
 
\subsection{Integration into a laboratory-based GISAXS setup}
\begin{figure}[!h]
     \centering
     \begin{subfigure}[b]{0.23\textwidth}
         \centering
         \includegraphics[width=\textwidth]{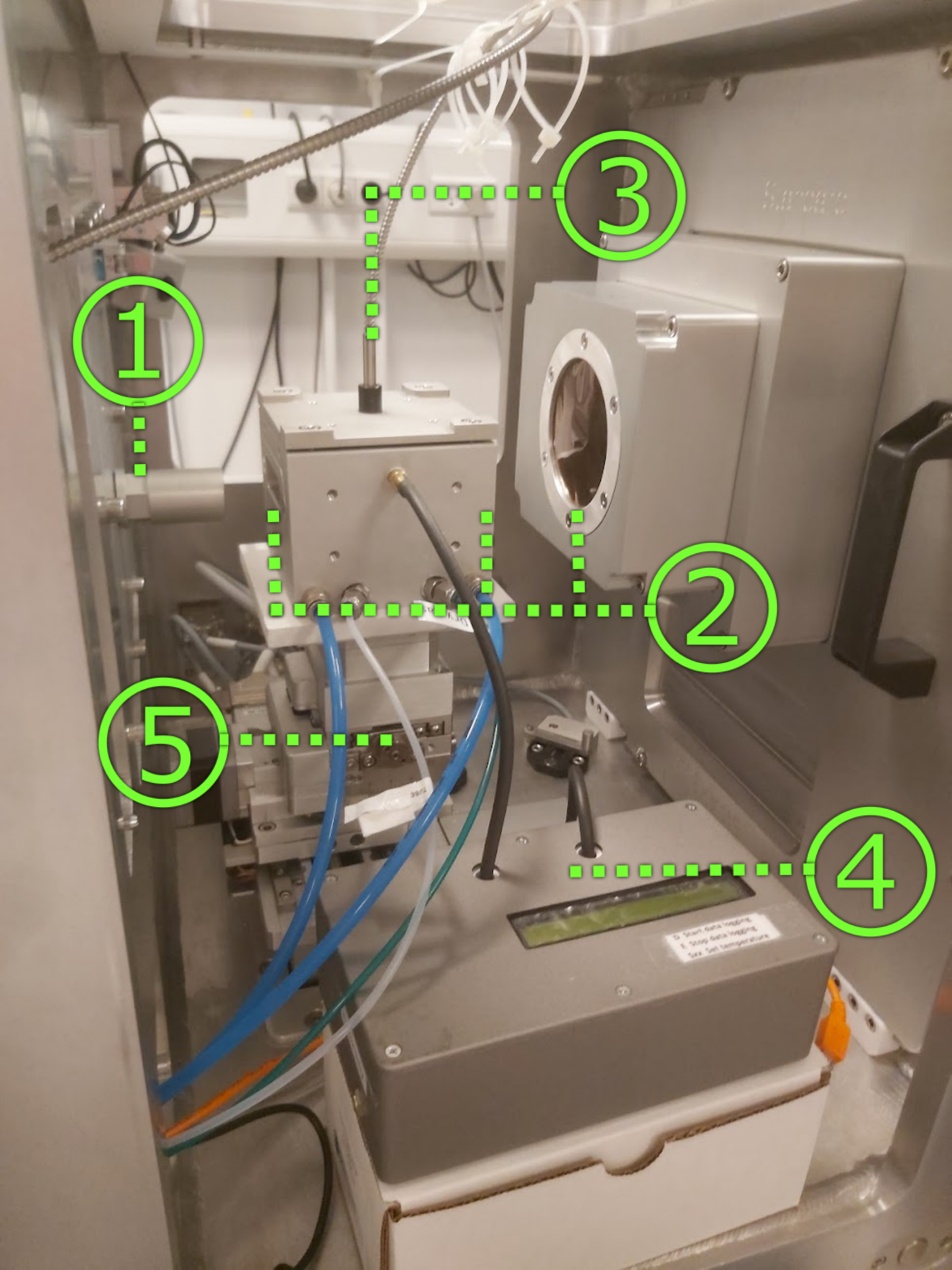}
         \caption{}
         \label{fig:enclosure}
     \end{subfigure}
     \hfill
     \begin{subfigure}[b]{0.23\textwidth}
         \centering
        \includegraphics[width=\textwidth]{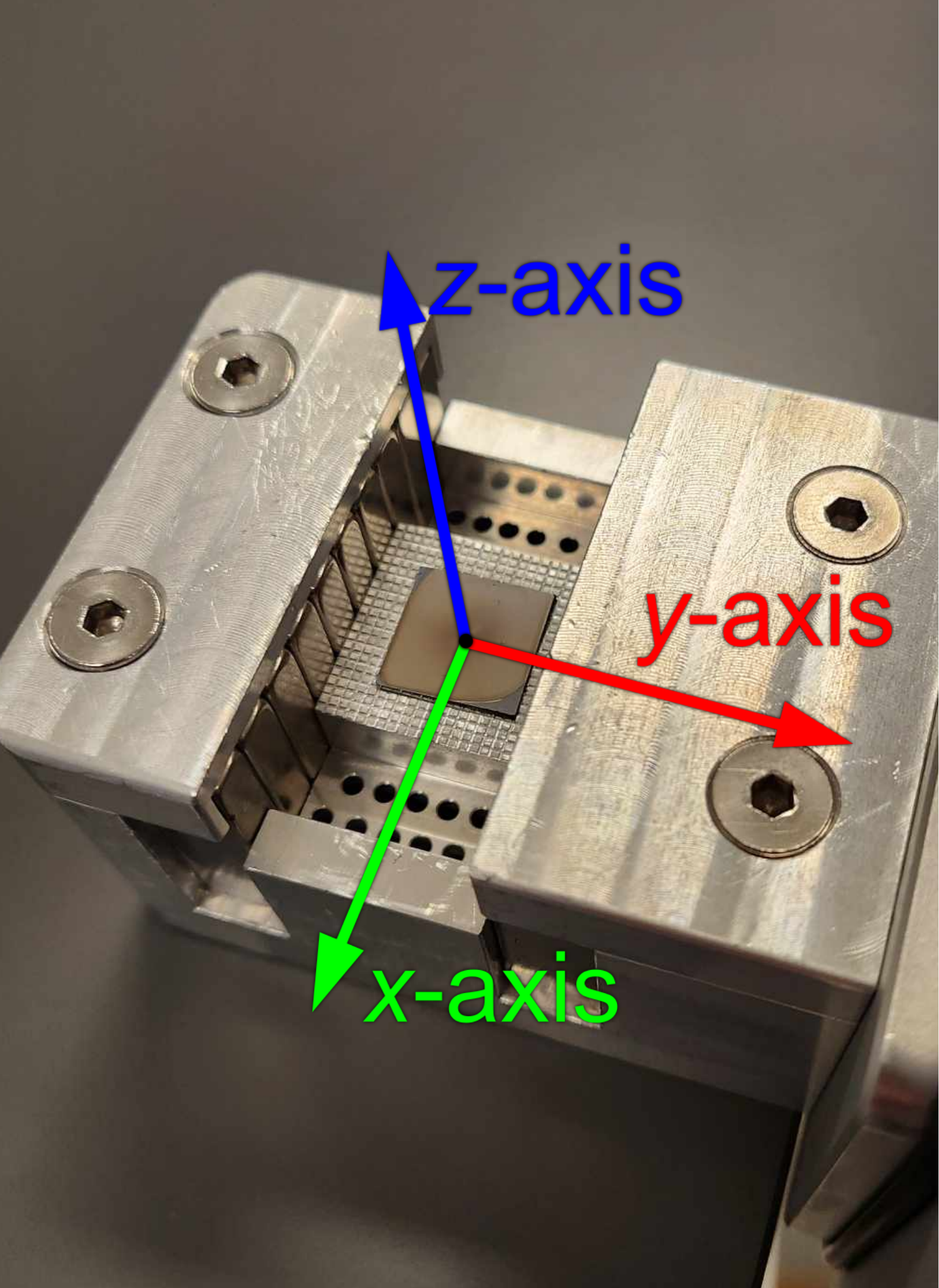} 
         \caption{}
         \label{fig:xyz}
     \end{subfigure}
        \caption{Photographs of the STVA chamber mounted inside the RUCSAXS sample‑environment enclosure. (a) Overview with key components labeled: (1) X‑ray beam from source, (2) Kapton windows in the beam path, (3) fiber-optic probe, (4) SVC unit, and (5) linear and rotary motion stage. (b) Sample mounted on the magnetic drawer stage, with axes annotated: the \textit{x} and \textit{y} directions lie in the sample plane, with \textit{x} along the beam direction, while \textit{z} is normal to the sample surface (out of plane).}
        \label{fig:integr} 
\end{figure}

At RUCSAXS (Roskilde University Interdisciplinary X-ray Scattering Hub), the STVA chamber is housed within the main sample environment enclosure of a XEUSS 3.0 UHR XL instrument (Xenocs, Grenoble, France), an advanced system designed for wide- and small-angle X-ray scattering experiments (see Fig. \ref{fig:enclosure}). The STVA chamber is mounted on a goniometer with two translational motors for sample positioning in the horizontal \textit{y}-direction and the vertical \textit{z}-direction, respectively (see Fig.~\ref{fig:xyz}). Rotational motors allow adjustment of the incident beam angle, $\alpha_i$, for GISAXS measurements and likewise, the sample can be tilted around the beam axis for alignment.

The CEM and MFC are positioned outside the main sample environment enclosure, with flow tubing directed into the STVA chamber through ports in the main enclosure  walls. Before exiting the main enclosure, the exhaust gas passes through the SVC unit before ultimately being discharged through the laboratory exhaust system. Water tubes between the STVA chamber and the Julabo circulator use the same main enclosure exit port, with tubing lengths kept minimal to reduce temperature gradients. Foam padding on the water pipes outside the main enclosure further mitigates temperature fluctuations. Experiments are performed at ambient pressure in the main enclosure which requires the detector tank vacuum to be separated from the main enclosure by cap with a Kapton window (thickness $\sim$75~$\mu$m) to allow passage of scattered X-rays (see Fig.~\ref{fig:enclosure}).

The reflectometer fiber-optic probe extends from the top of the STVA chamber and is  loosely mounted along the top of the enclosure leading to an exit port thus  minimizing strain on the optic fiber during sample alignment. The spectral reflectometer is positioned outside the main enclosure next to the CEM system.

Typical slit sizes defining the beam geometry for GISAXS are 0.04 mm vertically and 0.4 mm horizontally. In case of weakly scattering samples, this may be relaxed somewhat, e.g., 0.1 mm vertically and 0.5 mm horizontally, but with attention to the sample size and hence the maximal footprint on the sample. To start the GISAXS alignment, the sample stage is first translated in the horizontal ($x$) direction along the beam and then in the vertical ($z$) direction while the detector records the total photon signal (see Fig.~\ref{fig:xyz} for axis orientation). Sudden changes in the total number of counts reveal at which positions the beam enters and leaves the sample surface (footprint). In case of the RUCSAXS setup with an Eiger 2R 1M detector (Dectris, Baden, Switzerland), no beam stop is needed to protect the detector. With the sample located, height, $z$, and the incident angle, $\alpha_i$, are refined in an alternating procedure, while the two settings are strongly coupled. A scan in $z$ identifies the surface height at the half‐maximum intensity point, and, at that height, a scan of $\alpha_i$ finds the position where the direct‐beam intensity, as defined by a rectangular region of interest on the detector, is at its maximum. This angle is defined as the zero incoming angle to be fine-tuned in the following procedure. The $z$ and $\alpha_i$ settings are coupled and the scans are repeated twice more with progressively smaller step sizes and narrower ranges to fine-tune the alignment. Finally, $\alpha_i$ is raised to an angle between $0.1^{\circ}$ and $0.18^{\circ}$, a value which is below the critical angle for penetration of the (Si) substrate, but still large enough to provide a strong specular reflection separated from the direct beam on the detector.  Any mismatch between the observed and geometrically expected reflection position is checked with a custom-made plotting routine and corrected by an adjustment of $\alpha_i$ within $\pm 0.0025 \degree$. This uncertainty is linked to the estimate of the sample-to-detector distance and hence also to the size of the beam footprint on the sample, especially at low sample-to-detector distances. The procedure is typically executed for two different incoming angles, e.g., $0.11^{\circ}$ and $0.17^{\circ}$.

\section{Evaluation and Optimisation}\label{sec:Evaluation}
In this section, the operational parameters of the environmental chamber are assessed, including gas exchange times, magnetic field control, and performance during X-ray scattering experiments.

\subsection{Fill and quench of chamber}

\begin{figure}[!h]
     \centering
     \begin{subfigure}[b]{0.23\textwidth}
         \centering
         \includegraphics[width=\textwidth]{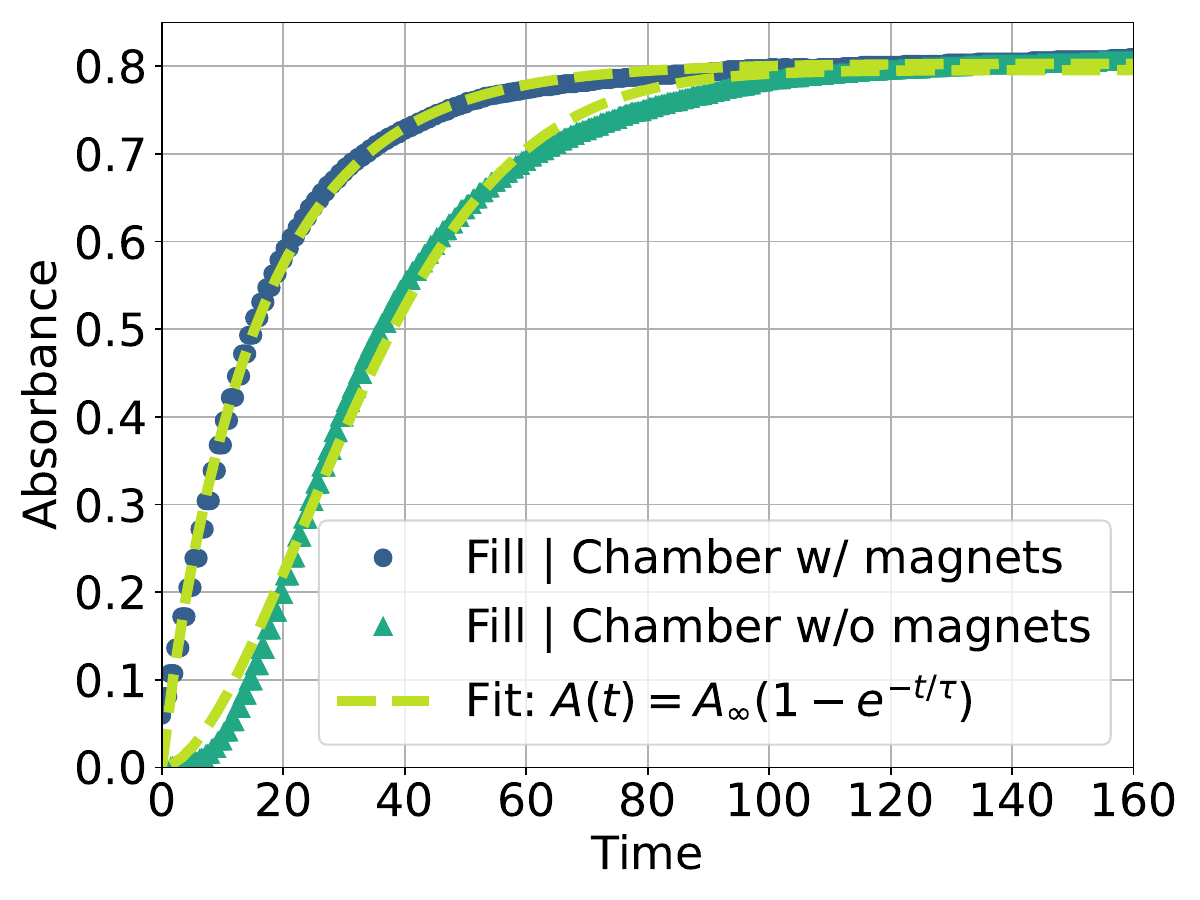}
         \caption{}
         \label{fig:filling}
     \end{subfigure}
     \hfill
     \begin{subfigure}[b]{0.23\textwidth}
         \centering
        \includegraphics[width=\textwidth]{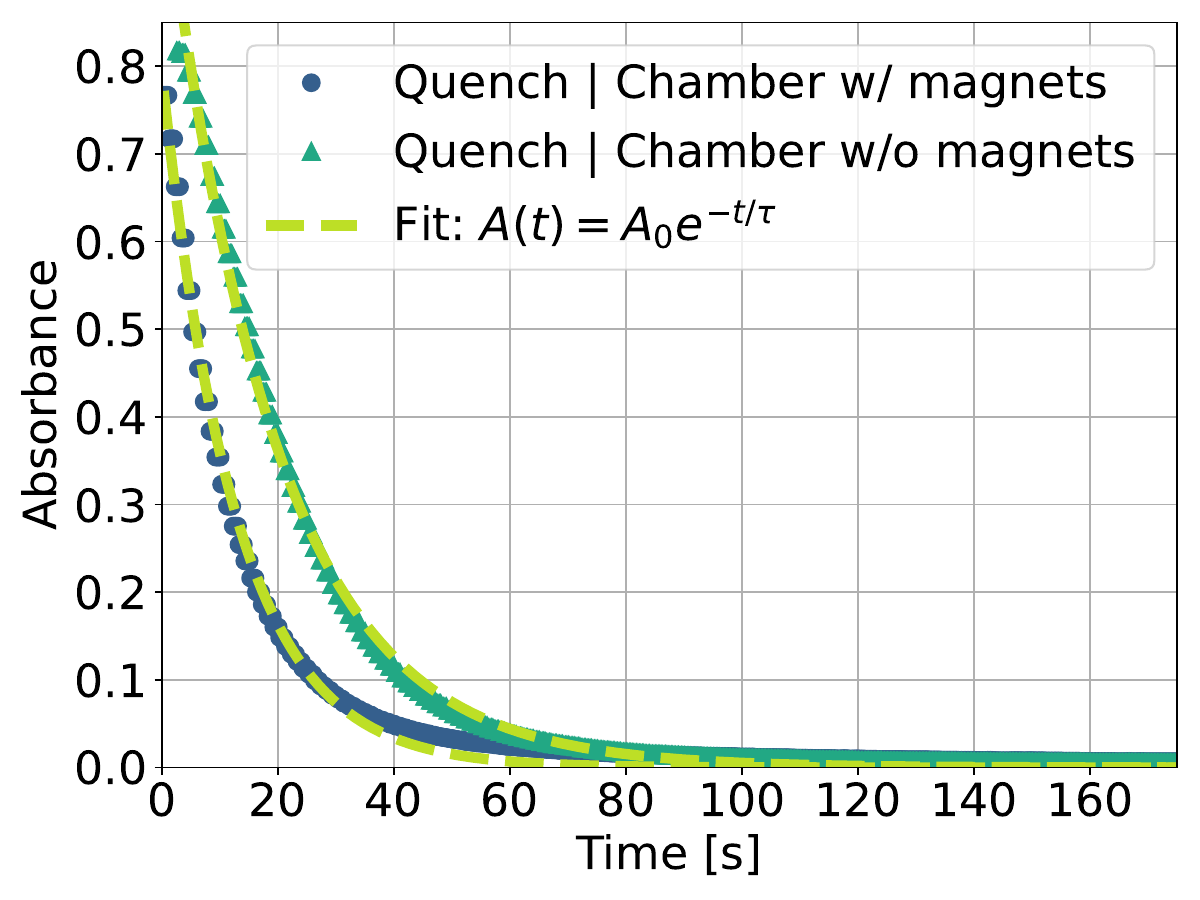}
         \caption{}
         \label{fig:quenching}
     \end{subfigure}
        \caption{(a) The absorbance as function of time during chamber filling with fits to eq.~\ref{A1} for both magnetic and base drawer. (b) The absorbance as function of time and fits to eq.~\ref{A2} during quenching. }
        \label{fig:fillquench}
\end{figure}

During annealing a certain protocol is followed where it is important to establish a new solvent vapor pressure in the annealing chamber as fast as possible. Likewise, in order to maintain a nanostructure obtained by solvent vapor annealing of block copolymers, fast drying (quench) times are often desirable\cite{Gotrik2013}. Because the characteristic filling or emptying time scales with the internal free volume, a smaller chamber is advantageous. The absorbance in the exhaust gas is proportional to the solvent-vapor concentration. Following  a step-change in inlet gas humidity during filling, this absorbance increases according to
\begin{equation}\label{A1}
A(t)=A_{\infty}\!\left[1-e^{-t/\tau}\right],    
\end{equation}
and for a quench with dry gas, the absorbance decays as
\begin{equation}\label{A2}
A(t)=A_{0}\,e^{-t/\tau},
\end{equation}
where $A(t)$ is the absorbance, $A_{\infty}$ is the final steady-state absorbance during filling reached after a long time, $A_0\equiv A(t=0)$ is the absorbance at the onset of a quench, and chamber volume, V, divided by gas flow, Q, $V/Q$, is the theoretical residence time constant, $\tau_{\mathrm{th}}$. The base drawer ($V=92$~cm$^3$, $Q=200$~sccm) gives $\tau_{\mathrm{th}} = 27.6$~s, while the magnetic drawer ($V=45$ cm$^3$, $Q=200$~sccm) gives 13.5~s (Tab.~\ref{table:fill}). During filling, the time required for $A(t)$ to change from 10\% to 90\% of $A_{\infty}$ is $\Delta t_{10\rightarrow 90}= \tau_{\mathrm{th}}\ln 9 \approx 2.20\,\tau_{\mathrm{th}}$.
We extract two characteristic times from each experimental absorbance curve in Fig.~\ref{fig:fillquench}: (i) a time constant from an exponential fit, $\tau_{\mathrm{fit}}$, and (ii) the 10–90 \% window $\Delta t_{10\rightarrow 90}=t_{90\%}-t_{10\%}$ (or $t_{10\%}-t_{90\%}$ for a quench), read directly from the curve.  The ratio $\Delta t_{10\rightarrow 90}/(2.20\,\tau_{\mathrm{fit}})$ stays within $\pm15\%$.

\begin{table}[!h]
\centering
\begin{tabular}{lll}
Parameter & Magnetic drawer & Base drawer \\
\hline
Free volume & 45~cm$^{3}$ & 92~cm$^{3}$ \\[2pt]
$\Delta t_{10\rightarrow 90}$ (200~sccm fill) & 40.9~s & 58.9~s \\
$\tau_{\mathrm{th}}$ (200~sccm fill) & 13.5~s & 27.6~s \\
$\tau_{\mathrm{fit}}$ (200~sccm fill) & 16.5~s & 25.6~s \\[4pt]
$\Delta t_{90\rightarrow 10}$ (600~sccm quench) & 27.3~s & 38.3~s \\
$\tau_{\mathrm{th}}$ (600~sccm quench) & 4.5~s & 9.2~s \\
$\tau_{\mathrm{fit}}$ (600~sccm quench) & 12.7~s & 19.0~s\\

\end{tabular}
\caption{Free volume and characteristic fill/quench times for magnetic drawer and base drawer. $\tau_{\mathrm{fit}}$ values are obtained from fitting exponential models to the experimental absorbance curves; $\Delta t_{10\rightarrow 90}$ and $\Delta t_{90\rightarrow 10}$ are read directly from the absorbance curves.}
\label{table:fill}
\end{table}

For the 200~sccm fill, $\tau_{\mathrm{fit}}$ differs from $\tau_{\mathrm{th}}$ by less than 20\%.  During the 600~sccm quench, however, $\tau_{\mathrm{fit}}$ is two–to–three times larger, likely because additional condensed solvent must evaporate from interior walls, tubing, and seals before the chamber is truly dry.  Compared with our previous setup\cite{Ariaee2023}, the characteristic filling and emptying times are reduced by a factor of 5–10, as expected based on the change in chamber volume.

\subsection{Application of a magnetic field during STVA}

\begin{figure}[!h]
    \centering
    \includegraphics[width=0.9\linewidth]{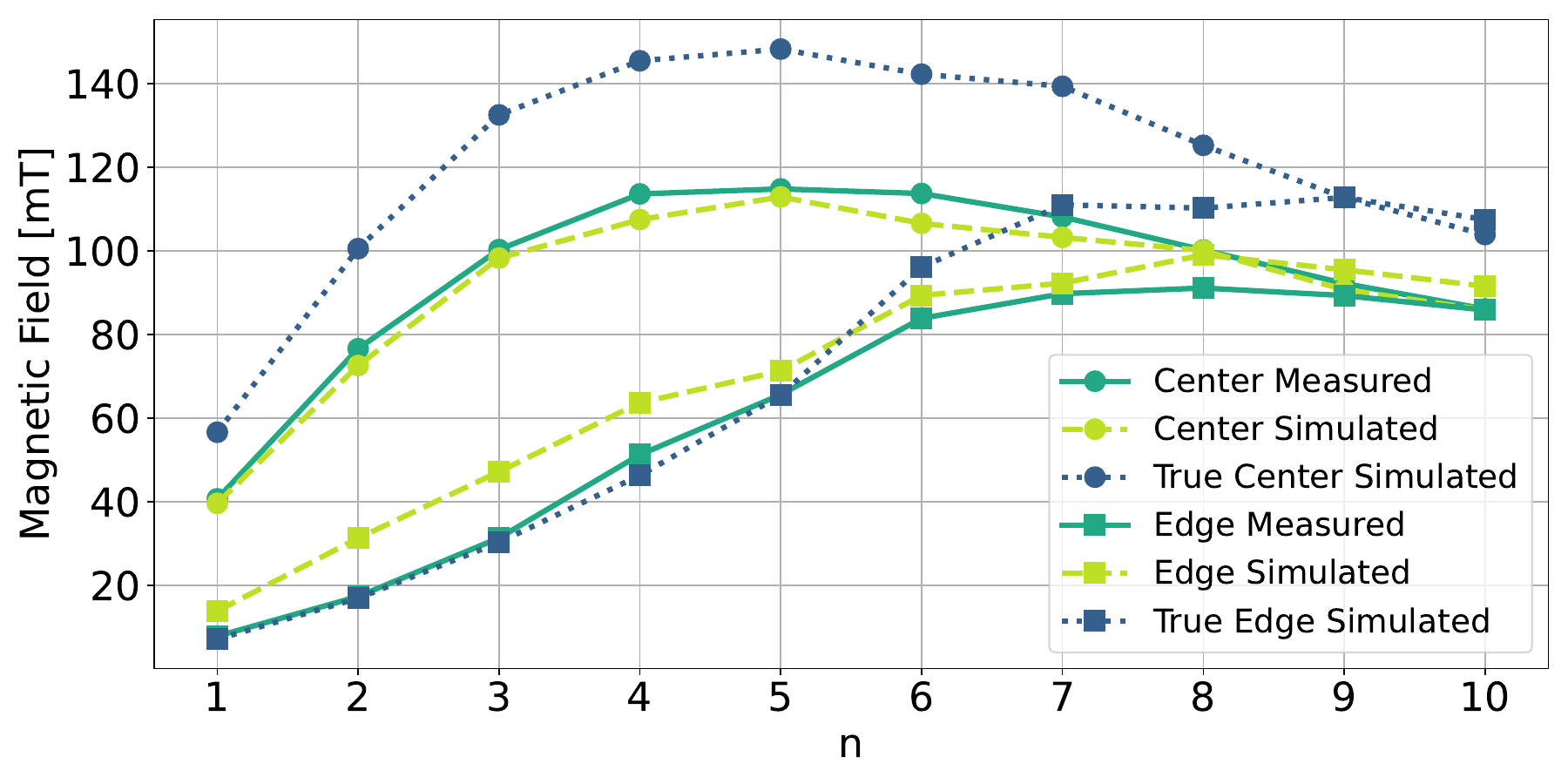}
    \caption{Magnetic field magnitude in mT measured with a Gauss probe (solid lines) and computed by FEM (dashed lines) at a height of $z=9$~mm above the sample stage, plotted versus the number of magnets (n) inserted on each side of the stage. Center values correspond to $(x,y)=(0,0)$ and edge values to $(x,y)=(\pm11~\mathrm{mm},0)$. Circle markers denote center data and square markers edge data.}
    \label{fig:nMagnets}
\end{figure}

The compact size of the STVA chamber, and therefore the size of the slot for inserting a sample table drawer, limits the available volume for the inclusion of permanent magnets: a $10 \times 30 \times 50$~mm$^3$ space on each side of the sample is available for this purpose. A magnetic field is generated by inserting up to 10  $10 \times 30 \times 5$~mm$^3$ N40 NdFeB permanent magnets, magnetized along the 5~mm axis, purchased from magnets4you GmbH. The magnets, up to 10 stacked together on either side of the sample table with their magnetization direction parallel to each other, can be aligned in different ways with the resulting field at the chamber table center either in the plane of the sample or parallel to the sample normal. In case of in-plane field (x-config., see Supplementary Material~Fig.~S2b), the field strength in the $x$-direction (along the X-ray beam path, cf.~Fig.~\ref{fig:xyz}) was measured by a Gauss meter 9~mm above the sample plane where the Hall element of the probe tip is located (cf. Supplementary Material~Fig.~S2a).
The measured field strengths above the sample center ($x$,$y$,$z$) = (0, 0, 9~mm) and edges ($\pm$11~mm, 0, 9~mm) are plotted as grass‑green markers in Fig.~\ref{fig:nMagnets} versus the number of magnets inserted. When the full $2\times10$ magnet array is in place, the two traces merge, indicating that the in-plane magnetic field achieves optimal homogeneity along the X-ray beam when the array is complete. Potential lateral variations in field strength along the \textit{y}-axis (perpendicular to the beam in the sample plane) were not investigated in this study.

The drawer is optimized for producing a field along the X‑ray beam path (x-config.), but the field can be directed out‑of‑plane along $z$ by rotating the magnetic bar arrays into a vertical orientation (rotation about the $x$‑axis). In this z-config. arrangement, using up to six magnets per side yields $B_z \approx 120$~mT at the stage center and about $100$~mT near the edges. Rotating the bar arrays by 90$^{\circ}$ within the sample (film) plane brings opposite poles face‑to‑face across the 22~mm gap between the magnets, so the field is redirected into y-config. and inherently amplified, along \textit{y}. Further increases are possible using thicker bar magnets, yet ultimately constrained by the drawer clearance of $10\times30\times50$~mm$^3$ around the sample. A schematic overview of the three orthogonal field geometries is available in Supplementary Material Fig.~S2b.

To further analyze the magnetic field distribution, we simulated the experimental setup using the package Elmer\cite{ElmerFEM}, a finite element method (FEM) solver. The simulation focused on a magnet array in air corresponding to the one used in the STVA setup, but neglecting the aluminum  STVA chamber and sample stage in order to simplify the model. The magnets were modeled as blocks with specified remanent flux density based on N40 NdFeB values from the vendors website\cite{MagnetShopN40}, $B_r =1.3~\mathrm{T}$. The surrounding air was modeled using the vacuum permeability,$\mu_{0}$. The results, presented in Fig.~\ref{fig:nMagnets} (light green), validate the experimental findings (grass-green), showing good agreement in field strengths 9~mm above the sample stage center and the edges for all numbers of magnets inserted, though with a notably larger maximum relative deviation at the edges (80\%) than at the center (6\%) . The FEM analysis also revealed the field distribution beyond the measurement range of the Gauss probe, namely at the level of the sample position at ($x$, $y$, $z$) = (0, 0, 0 mm) and ($x$, $y$, $z$) = ($\pm$11~mm, 0, 0~mm) (dark-blue curves). 

 The full magnetic field landscape at the sample stage position as simulated with FEM is shown in Supplementary Material Fig.~S2c. From this analysis we estimate the in-plane field within an area of 20 x 20 mm$^2$ of the sample stage (corresponding to a typical sample size) to be slightly higher than 100~mT with all 2 x 10 permanent magnets inserted.

\subsection{Stability of reflected X-ray beam}

When testing \textit{in situ} GISAXS during STVA of polymer thin films, the position of the reflected beam on the detector was observed to show large variations during use of large relative solvent humidities (>70\%) at room temperature. By placing the sample on a sanded aluminum plate of a size similar to the sample or by placing the sample on 3 small shards of Si wafer, the reflected beam position became stable. We conclude that the original smoothness of the sample stage allows for solvent to become trapped between sample and sample table via capillary forces resulting in a change of the  angle of incidence and hence a change in exit angle of the reflected beam. As the solvent humidity in the chamber changes, so does the amount of solvent trapped and hence the position of the reflected beam on the detector varies with time in an unpredictable and non-reproducible way. The final sample table has thus been machined with a square pattern featuring a periodicity of 1~mm (cf. Supplementary Material Fig.~S3), which has successfully eliminated problems with unstable beam angles relative to the sample surface.
\FloatBarrier

\section{Research examples}\label{RandD}
We present four examples highlighting the versatility of the STVA setup for thin film studies. The examples span magnetic field-driven nanoparticle assembly, morphological transformations in BCPs under controlled solvent atmosphere and at different temperatures and a demonstration of \textit{in situ} GISAXS with a lab-based instrument. All polymers used were purchased from Polymer Source Inc. (Quebec, Canada); all polymer solutions were passed through 2~$\mu$m PTFE filters and used without further purification. Single-side-polished Si(100) wafers from Topsil GlobalWafers (Frederikssund, Denmark) served as substrates and were used as received unless stated otherwise. Polymer thin films were made by spin coating polymer solutions in toluene onto Si substrates at 3000~rpm with an acceleration of 4000~rpm~s$^{-1}$.

\subsection{Magnetic field-driven nanoparticle assembly in a block copolymer thin film template during STVA}
The setup allows STVA under the application of a magnetic field as demonstrated with a system of magnetic $\gamma$-Fe$_2$O$_3$ nanoparticles dispersed in a block coplymer thin film during solution coating. The magnetic nanoparticles are coated with a thin polymer layer with the aim of rendering the particles preferentially miscible with one of the polymers in the BCP \cite{Hartmann2023,Konefa2021,Zhu2018}. In this particular experiment, the BCP is a polystyrene‑\textit{b}‑polybutadiene diblock copolymer (PS‑\textit{b}‑PB) while the magnetic nanoparticles are coated with PS. PS‑\textit{b}‑PB has distinct mechanical properties at RT, where the minority block PS is in the glassy state  (the glass transition temperature of PS is around 100~$^{\circ}$C) and expected to be forming PS cylinders\cite{Bates1990} in a matrix of the majority block, PB, which is in the rubber state (glass transition temperature of PB is typically around -85~$^{\circ}$C). The $\gamma$-Fe$_2$O$_3$ particles are synthesized following the procedure in Ref.~\onlinecite{Cabuil1994} and measured by transmission electron microscopy to have a diameter of $10.2 \pm 2.7$~nm prior to coating with PS (cf. Supplementary Material Figs.~in~S4).
To prepare the hybrid film, we first prepare a mixture of 100~$\mu$l 3~wt\% PS(89.6)-\textit{b}-PB(226), where the numbers in parentheses denote the number-average molecular weight $M_n$ in kg~mol$^{-1}$, with a polydispersity (PDI) of 1.05, and 50~$\mu$l PS-coated 1~wt\% $\gamma$-Fe$_2$O$_3$ nanoparticles, both dispersed in toluene. Solution coating is performed by pipetting 40~$\mu$l of the mixture onto a Si wafer ($15 \times 15$~mm$^2$), situated on the sample stage of the drawer with 10 magnets inserted on either side in the parallel magnetic field configuration. The drawer is slotted into the STVA chamber and a flow of 200~sccm 80\% or nominally 100\% toluene-saturated nitrogen (in praxis slightly undersaturated) is established. The film is left to dry in the toluene atmosphere at 24~$^{\circ}$C for at least 2~h after which the chamber is flushed with 600~sccm dry nitrogen.

\begin{figure}[!h]
    \centering
    \includegraphics[width=0.9\linewidth]{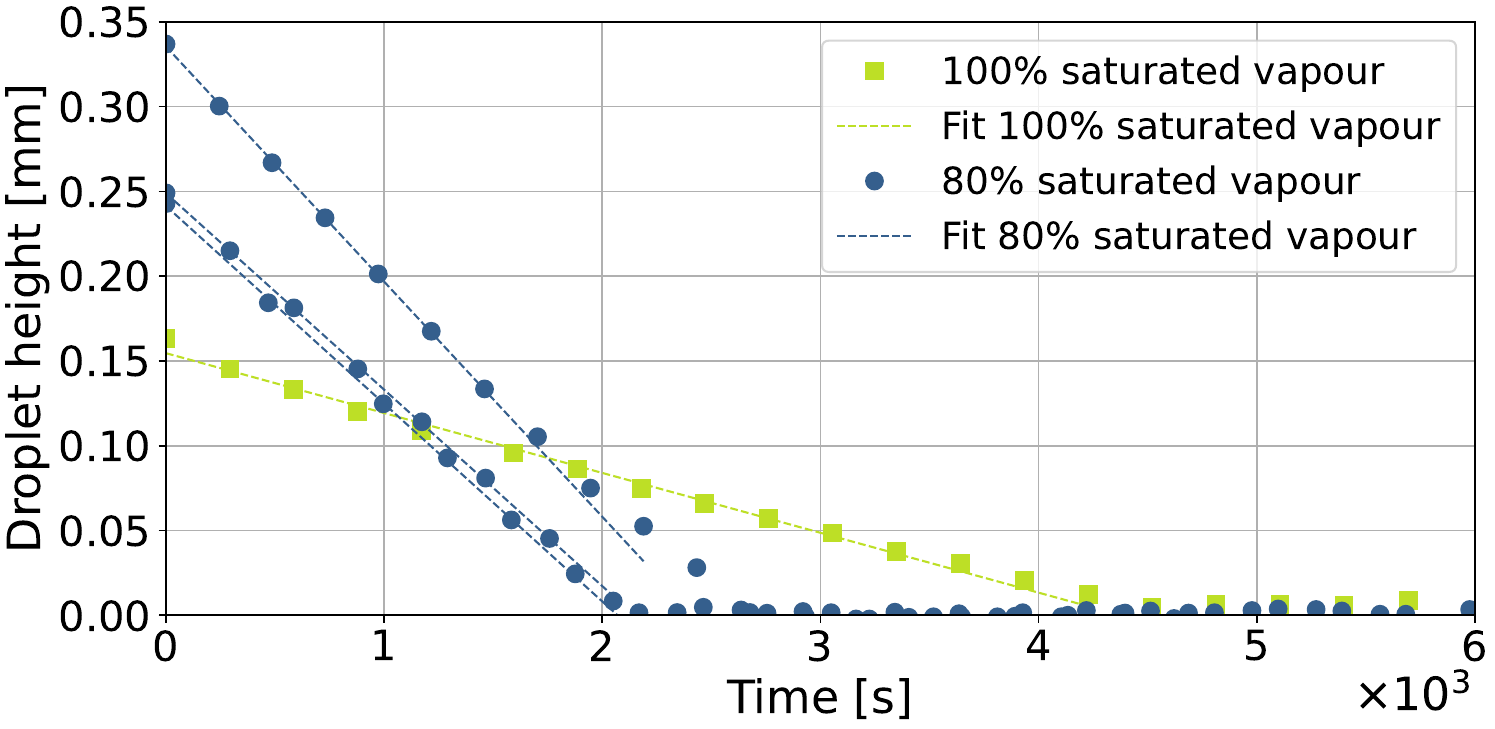}
    \caption{Evaporation rates of BCP/nanoparticle solution droplets left to dry in the STVA chamber with 80\% or nominally 100\% toluene-saturated nitrogen flow, as recorded from the halfway point of an X-ray scan along the droplet normal. Also shown are linear fits (dashed lines) to the data (filled dots or squares)}
    \label{fig:z_dial}
\end{figure}

The progress of the drying of the droplet is tracked by mounting the STVA chamber in the RUCSAXS experimental chamber and using the X-ray beam to scan along the height of the droplet. The position of the droplet-gas interface is found as the halfway point of the X-ray transmission curve. Fig.~\ref{fig:z_dial} shows how the 3 samples that were exposed to a lower concentration of toluene gas had a significantly increased drying rate averaging at 0.45~mm/h compared to the sample exposed to the highest possible concentration of toluene gas during drying where the evaporation rate was 0.13~mm/h. The drying rates are found by linear fits to the experimental data as shown in Fig.~\ref{fig:z_dial}.

\begin{figure}[!h]
     \centering
     \begin{subfigure}[b]{0.23\textwidth}
         \centering
         \includegraphics[width=\textwidth]{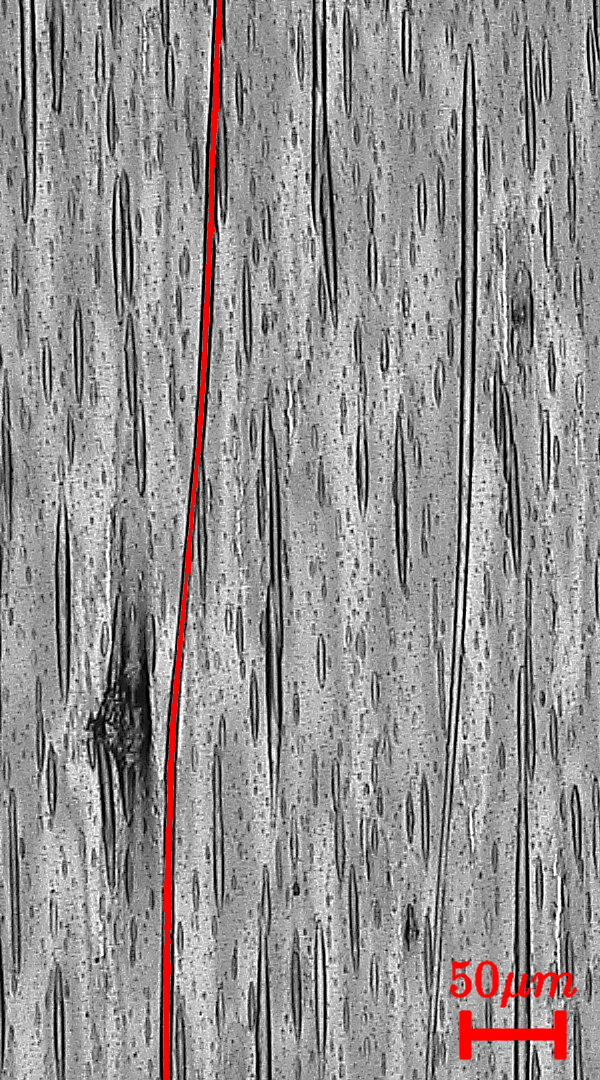}
         \caption{}
         \label{fig:80p}
     \end{subfigure}
     \hfill
     \begin{subfigure}[b]{0.23\textwidth}
         \centering
        \includegraphics[width=\textwidth]{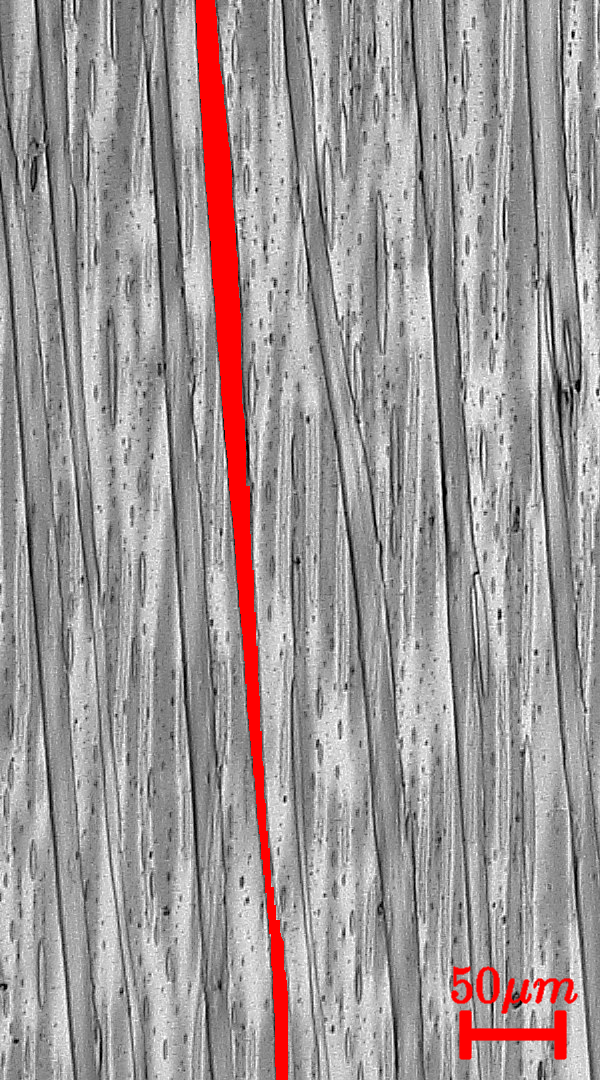}
         \caption{}
         \label{fig:100p}
     \end{subfigure}
        \caption{Optical microscopy images of linear nanoparticle aggregates formed with an applied parallel magnetic field during drying of a drop of toluene with a mixture of PS(89.6)-\textit{b}-PB(226) and magnetic PS-coated $\gamma$-Fe$_2$O$_3$ nanoparticles with a toluene relative humidity of (a) 80\% and (b) nominally 100\% in the nitrogen gas flow through the STVA chamber. One wire has been highlighted in red for each case, showing that slower solvent evaporation leads to thicker, more densely packed wires.} 
        \label{fig:micro}
\end{figure}

Optical microscopy images from the dried samples are shown in Figs.~\ref{fig:80p} and \ref{fig:100p}. The images show clear wire-like structures formed along the direction of the applied field. Furthermore, comparison between samples formed with different rates of drying, reveal that slower drying allow the nanoparticles more time to form longer and thicker nanoparticle aggregates. The wire widths in Fig.~\ref{fig:micro} are in the few-micron regime, and for the nominally 100\% toluene humidity case produces a higher density of wires >1~mm in length, while the 80\% case more frequently forms shorter, oblong aggregates at this resolution. Large-area microscopy images are provided in the Supplementary Material (Figs.~S5a,b), together with an image of a control sample prepared without application of a magnetic field during drying showing no apparent structure formation (Supplementary Material Fig.~S5c). Supplementary Material Figs.~S5d,e show analogous drying experiments performed on PS and PB homopolymer films with the same nanoparticle loading. In PS, the nanoparticles show only very limited formation of linear assemblies, whereas the PB film exhibits pronounced cracking and film delamination, consistent with PS-coated nanoparticles being incompatible with PB and thus no longer remaining embedded in the PB matrix.

\begin{figure}[H]
    \centering
    \begin{subfigure}[b]{0.45\textwidth} 
        \centering
        \includegraphics[width=\textwidth]{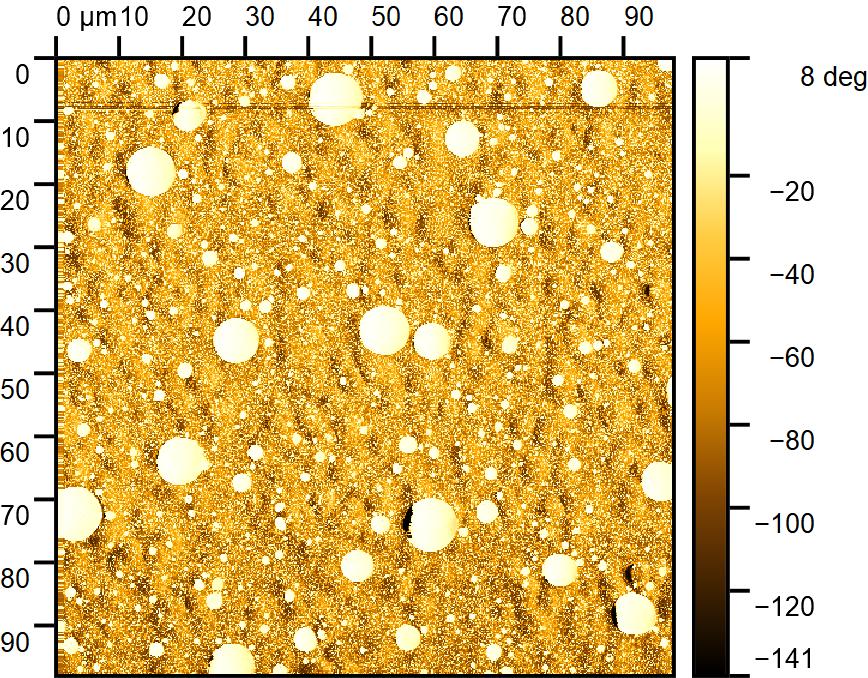}
        \caption{}
        \label{fig:p-afm1}
    \end{subfigure}
    
    \vspace{0.5cm} 
    
    \begin{subfigure}[b]{0.45\textwidth}
        \centering
        \includegraphics[width=\textwidth]{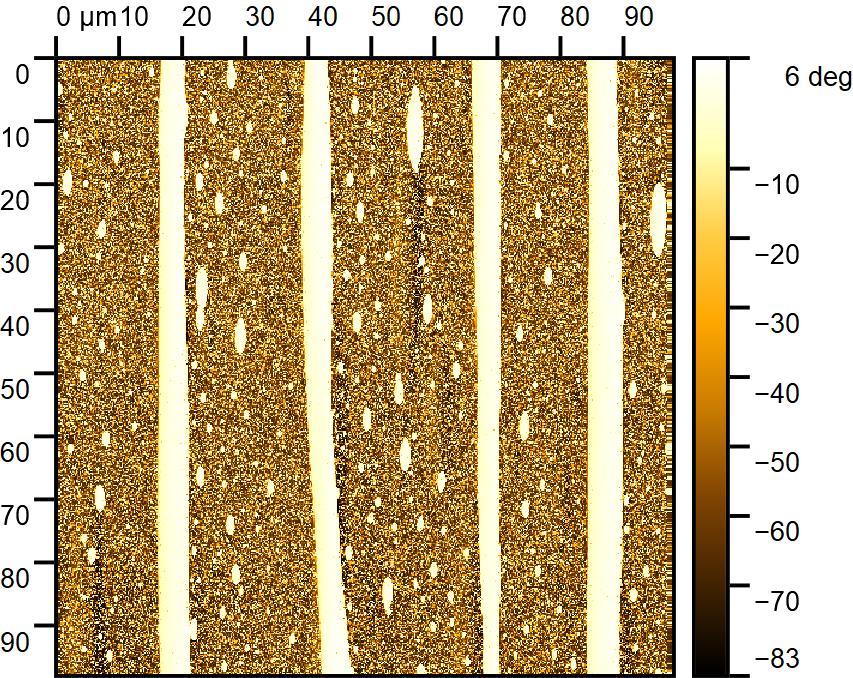}
        \caption{}
        \label{fig:p-afm2}
    \end{subfigure}
    
    \caption{Phase contrast AFM images after drying a drop of toluene with a mixture of PS(89.6)-\textit{b}-PB(226) and magnetic $\gamma$-Fe$_2$O$_3$ nanoparticles in an atmosphere of nitrogen saturated with toluene: (a) with no magnetic field applied during drying, showing a random nanoparticle aggregate distribution; and (b) with $\sim$ 100 mT parallel magnetic field applied during drying, showing continuous linear assemblies, aligned in the direction of the applied field. In both images, the lighter, larger features correspond to aggregated nanoparticles, whereas the darker, speckled regions correspond to the underlying polymer matrix.}
    \label{fig:p-afm}
\end{figure}

AFM-phase images of films dried in an atmosphere of nominally 100\% toluenen humidity with and without a parallel magnetic field applied during drying are shown in Fig.~\ref{fig:p-afm}. The glassy and rubbery phases in the BCP yields high contrast in AFM-phase imaging. Hence, from Fig.~\ref{fig:p-afm1} it is evident that when no field is applied, the nanoparticle/BCP mixture has no tendency to form ordered elongated structures, while Fig.~\ref{fig:p-afm2} demonstrates that such linear assemblies are formed when a magnetic field is applied during drying.

This behavior is consistent with the PS-coated nanoparticles preferentially partitioning into PS domains during solvent removal and experiencing dipolar interactions which with application of the parallel magnetic field favor end-to-end chaining. As these chains grow they exceed the spacing between neighboring PS microdomains (cylinder center-to-center spacing $\approx$ 130 nm, determined from GISAXS (not shown)), enabling aggregates to bridge multiple domains and coalesce into continuous linear assemblies, in agreement with the mechanism reported by Yao~et~al.\cite{Yao2014}.

This experiment confirms that our STVA chamber allows for control of drying rates and promotes reorganization of nanoparticles embedded in a BCP thin film with an applied magnetic field during drying. In this specific case, PS-coated $\gamma$-Fe$_2$O$_3$ nanoparticles self-assemble into wire-like structures growing from cylinder microdomains in the PS‑\textit{b}‑PB BCP matrix.

\subsection{STVA and \textit{ex situ} GISAXS of a di-BCP thin film}

\begin{figure}[H]
     \centering
     \begin{subfigure}[b]{0.45\textwidth}
         \centering
         \includegraphics[width=\textwidth]{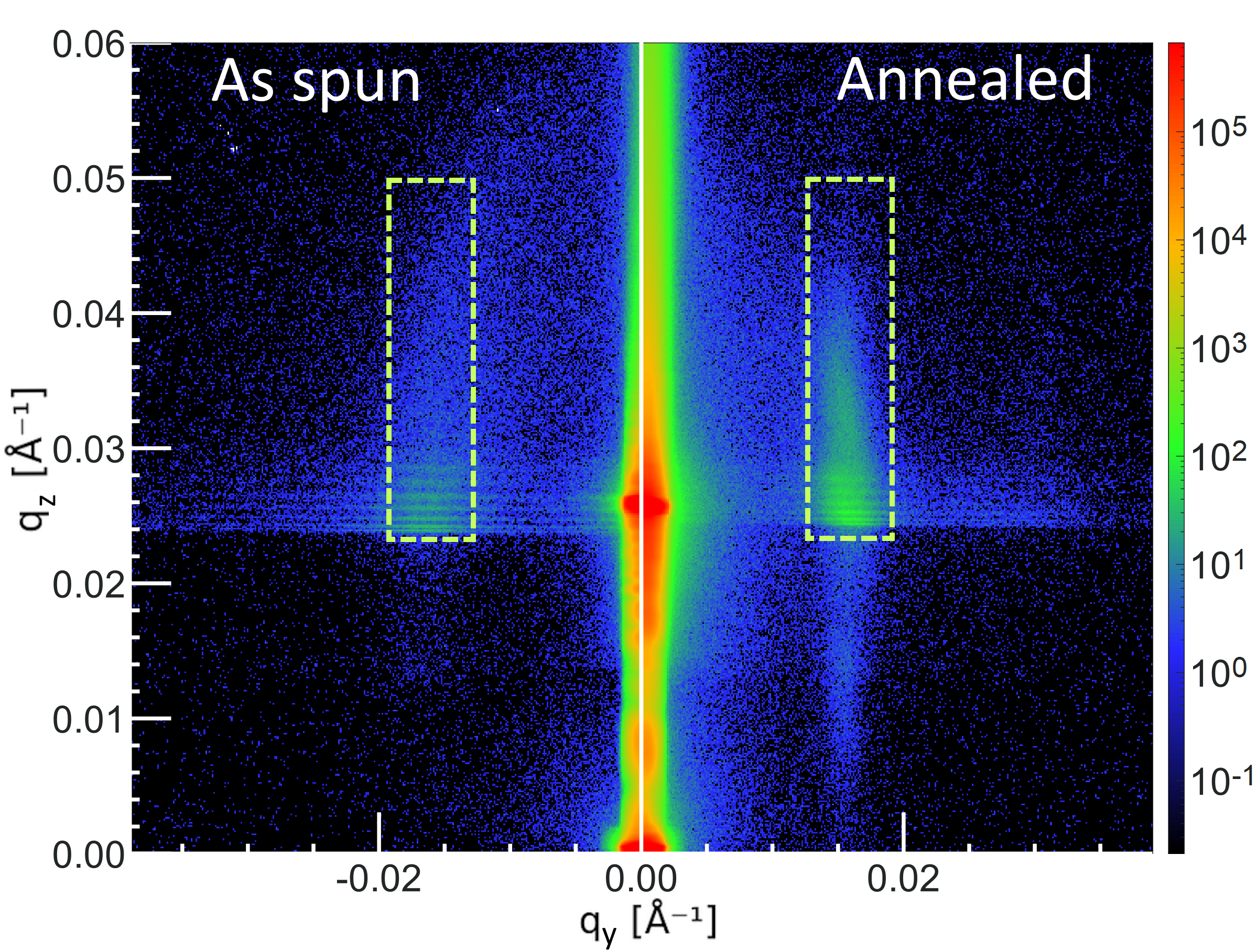}
         \caption{}
         \label{fig:PSPB60GISAXS}
     \end{subfigure}
     
     \vspace{0.5cm} 
     
     \begin{subfigure}[b]{0.35\textwidth}
         \centering
         \includegraphics[width=\textwidth]{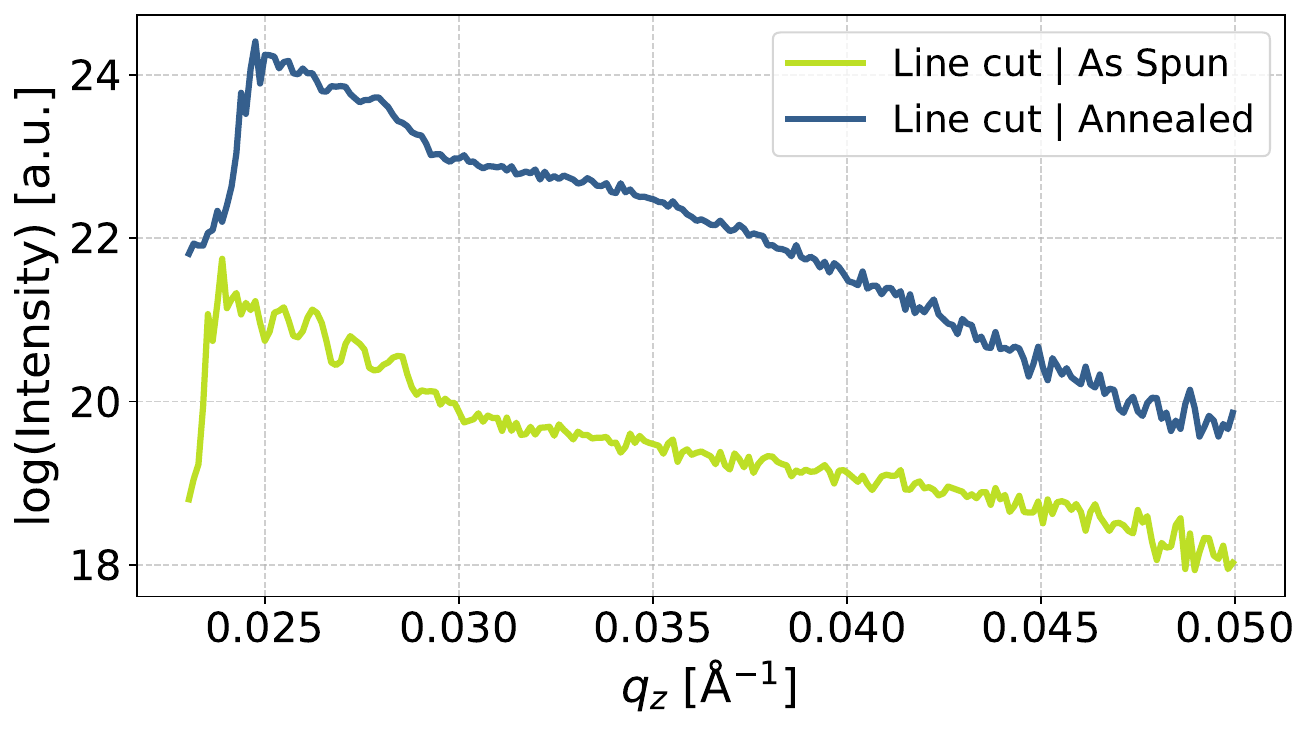}
         \caption{}
         \label{fig:PSPB60GISAXS_plot}
     \end{subfigure}

     \begin{subfigure}[b]{0.35\textwidth}
         \centering
         \includegraphics[width=\textwidth]{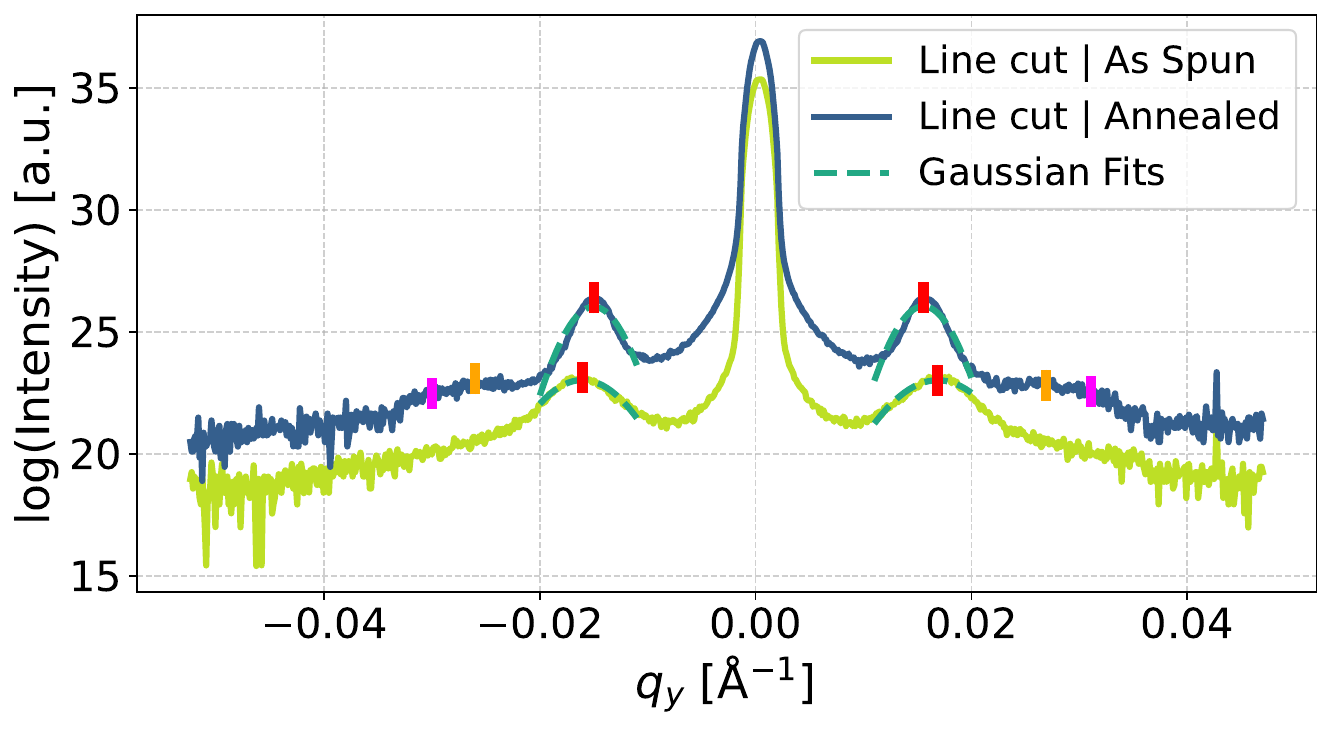}
         \caption{}
         \label{fig:PSPB60GISAXS_plot2}
     \end{subfigure}
     
    \caption{(a) GISAXS maps of PS(21.7)-\textit{b}-PB(60) thin film: unannealed (as-spun) (left) and after 4~h of solvent annealing in a 80$\%$ toluene humidity atmosphere  (right). The vertical boxed areas indicate the line cuts used in (b). The color scale is in  arbitrary units. (b) Summed intensity within the marked vertical box regions plotted along $q_z$, with curves vertically offset for clarity. Here the shoulder present above 0.03~Å$^{-1}$ for the annealed film indicates an emerging 3-dimensional ordered structure. (c) Horizontal line cuts at the $q_z$ position corresponding to the Yoneda region. The position of first-order peaks are found by fitting a Gaussian profile. First-order peak positions from these fits are marked by red ticks. Higher-order peak position as expected for hexagonal packing of lying cylinders are marked at $\sqrt{3}q_1$ and $2q_1$ with orange and magenta ticks, respectively.}
     \label{fig:PSPB60GISAXS_Highlight}
\end{figure}

A 3~wt\% PS(21.7)-b-PB(60) di-BCP in toluene was spin-coated onto a Si wafer and subjected to 4~h of room-temperature solvent vapor annealing in the STVA chamber with 80\% toluene humidity in the nitrogen flow through the chamber. The \textit{ex situ} GISAXS map was measured for 15000~s before and after annealing  with an incidence angle of 0.18$^{\circ}$ (in the dynamical regime between di-BCP and substrate critical angles), and is shown in Fig.~\ref{fig:PSPB60GISAXS}. From the relative size of the two polymer blocks, the BCP is expected to show an ordered structure with PS cylinders in a PB matrix\cite{Bates1990}. Fig.~\ref{fig:PSPB60GISAXS} shows one half of each GISAXS map put together. The GISAXS half-map to the left shows a Debye-Scherrer ring indicating randomly oriented cylinders before annealing. The GISAXS half-map to the right indicate the emergence of a 3-dimensional ordered structure of lying cylinders after STVA by the disappearence of the Debye-Scherrer ring and the appearance of an elongated Bragg spot. These observations are emphasized by the presence of a shoulder after STVA in the line cuts along $q_z$ at the $q_y$ position of the lateral peak in the Yoneda band\cite{MllerBuschbaum2009}  (the areas in the GISAXS maps used for making these line cuts are marked by rectangles in Fig.~\ref{fig:PSPB60GISAXS_plot}). Line cuts along $q_y$ at the $q_z$ position of the Yoneda band (corresponding to the horizontal region marked for another sample in Fig.~\ref{fig:GISAXSmask}) show a broad feature after STVA extending in $q_y$ from $\sqrt{3}q_1$ to $2q_1$ (these positions are marked with tics in Fig.~\ref{fig:PSPB60GISAXS_plot2}), where $q_1$ is the first-order lateral structural peak, the position of which is found by fitting a Gaussian profile. The occurrence and position of the broad higher-order features are consistent with a hexagonal ordering of the lying cylinders. For a highly ordered structure, we would expect to see peaks rather than shoulders at these positions\cite{Berezkin2018}. The fact that only broad shoulders and not distinct Bragg peaks are observed suggests that long‑range order is still limited in the film.

\begin{figure}[!h]
    \centering
    \begin{subfigure}[b]{0.45\textwidth} 
        \centering
        \includegraphics[width=\textwidth]{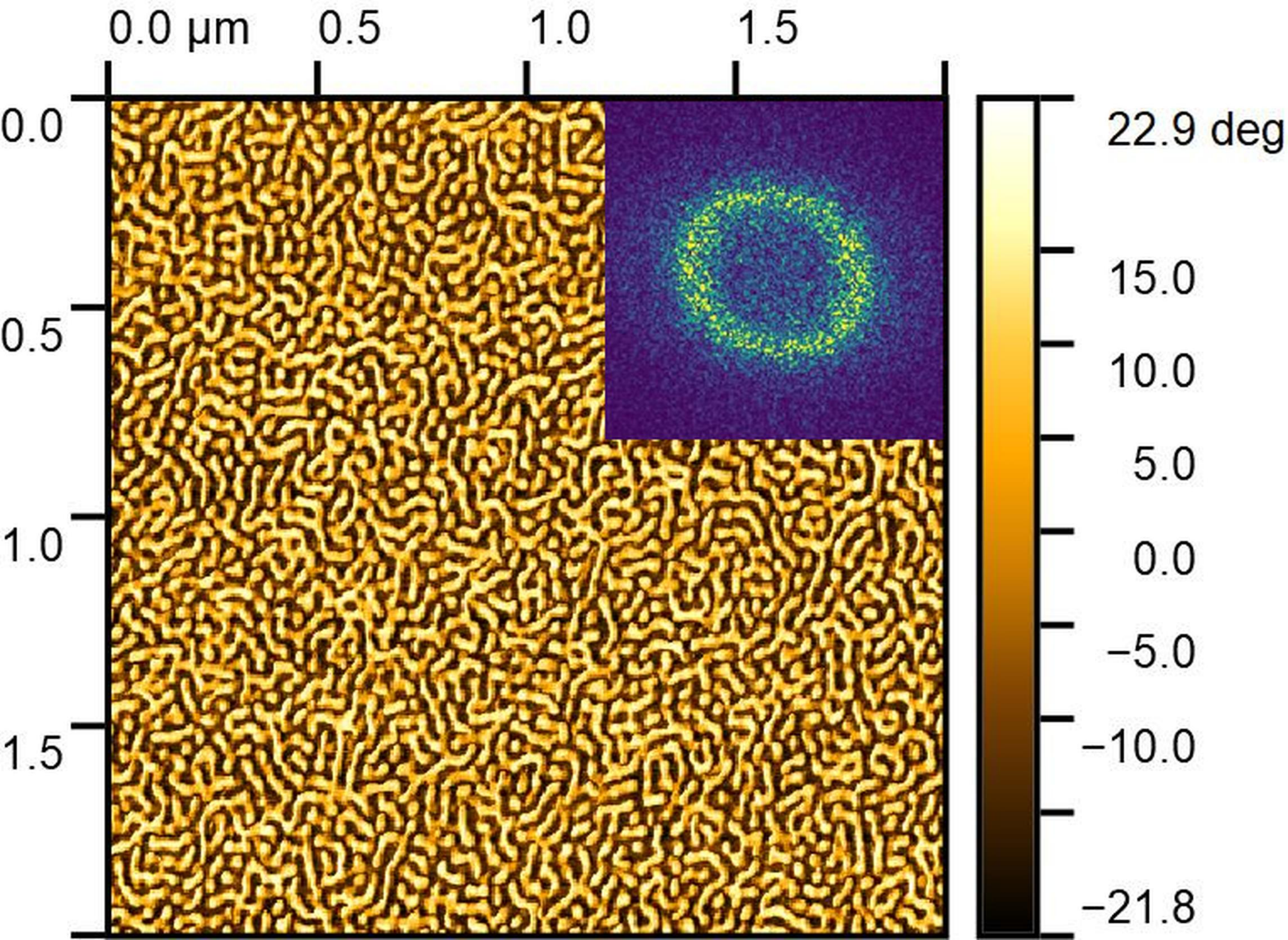}
        \caption{}
        \label{fig:p-afm3}
    \end{subfigure}
    
    \vspace{0.5cm} 
    
    \begin{subfigure}[b]{0.45\textwidth}
        \centering
        \includegraphics[width=\textwidth]{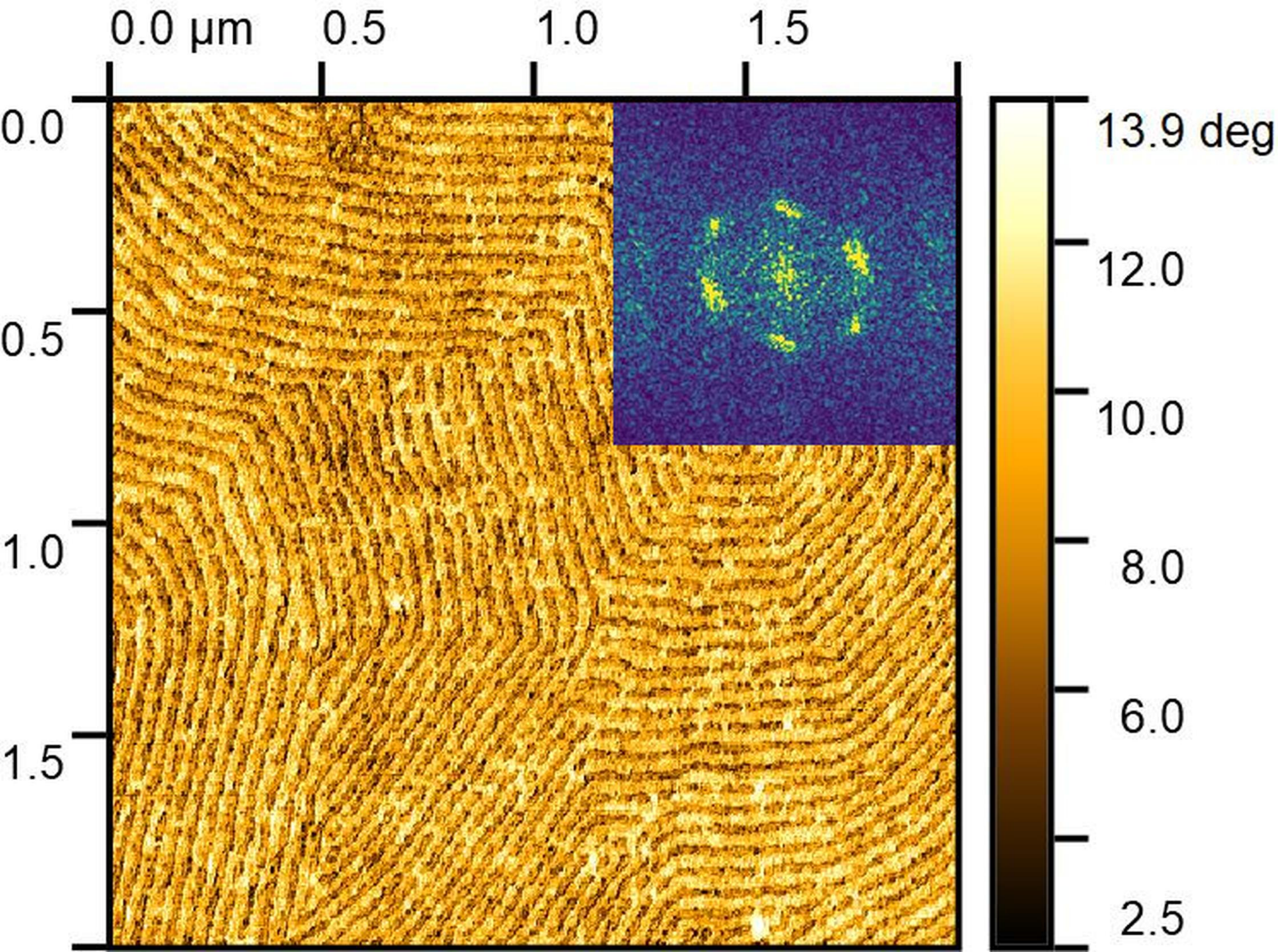}
        \caption{}
        \label{fig:p-afm4}
    \end{subfigure}
    
    \caption{AFM phase images and their Fourier transforms for PS(21.7)-\textit{b}-PB(60) thin films: (a) As-spun film, showing lying cylinders with random in-plane orientations. (b) After 4~h of solvent annealing, where grains of lying cylinders have aligned along preferred in-plane directions. Although the cylinders themselves remain parallel to the substrate, the selection of discrete grain orientations produces an averaged Fourier transform (see inset) with hexagonal symmetry. Cf.~Fig.~S7a for full‐size maps.}
    \label{fig:p-afmPSPB60}
\end{figure}

Phase imaging AFM is consistent with randomly oriented cylinders before STVA (Fig.~\ref{fig:p-afm3}), while large unidirectional grains are seen to have formed after solvent annealing (Fig.~\ref{fig:p-afm4}). The Fourier transform of the AFM image of the annealed film indicates preferred domain orientations with hexagonal symmetry in the film surface plane. The same sixfold symmetry is retained in a much larger field-of-view documented in the Supplementary material Fig.~S7a, where a minority of standing cylinders near the film surface also are evident. The 'dry' repeat distance has increased after annealing in a 80\% toluene humidity atmospshere, from 38.2$\pm0.1$ to 41.1$\pm0.1$~nm as found from GISAXS maps, and from 41.3 to 44.6~nm found using AFM as shown in Fig.~S7b. Both AFM and GISAXS show that annealing in the present case results in an increase of the 'dry' microdomain periodicity which is not uncommon \cite{Zhang2014, Gunkel2015} - however, as seen in research example C, the opposite may also be the case. Likewise, it is not surprising that AFM and GISAXS give different values for the microdomain periodicity. In the present case, the GISAXS values which are an average over the whole film depth, are 5-10$\%$ lower than the AFM values, which are the microdomain periodicitiy in the outermost surface layer, where 'reconstruction' due to the polymer-air interface may occur.

\subsection{\textit{In situ} GISAXS on a laboratory instrument during STVA of a BCP thin film} 
\begin{figure}[!h]
     \centering
     \begin{subfigure}[b]{0.45\textwidth}
         \centering
         \includegraphics[width=\textwidth]{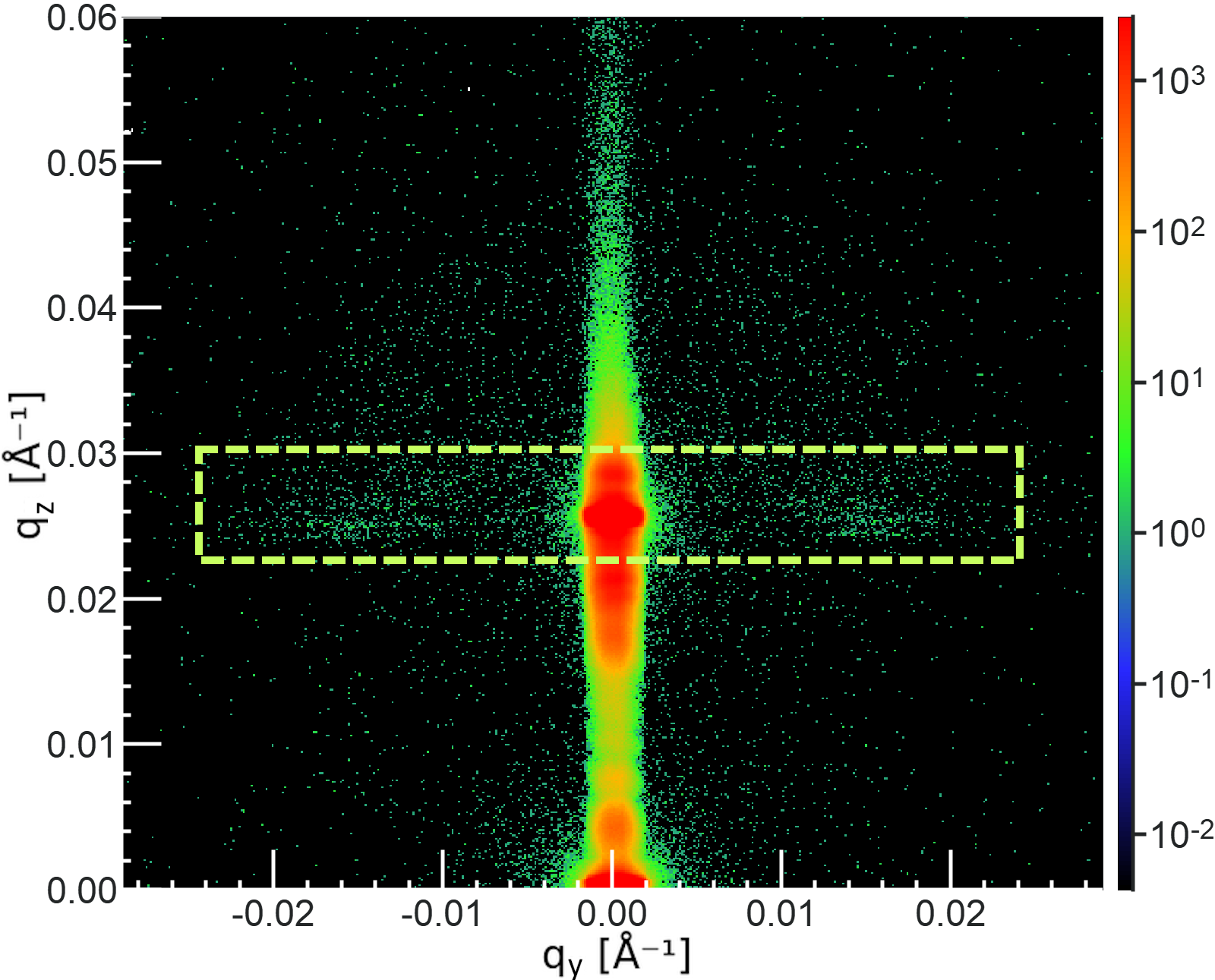}
         \caption{}
         \label{fig:GISAXSmask}
     \end{subfigure}
     
     
     \begin{subfigure}[b]{0.45\textwidth}
         \centering
         \includegraphics[width=\textwidth]{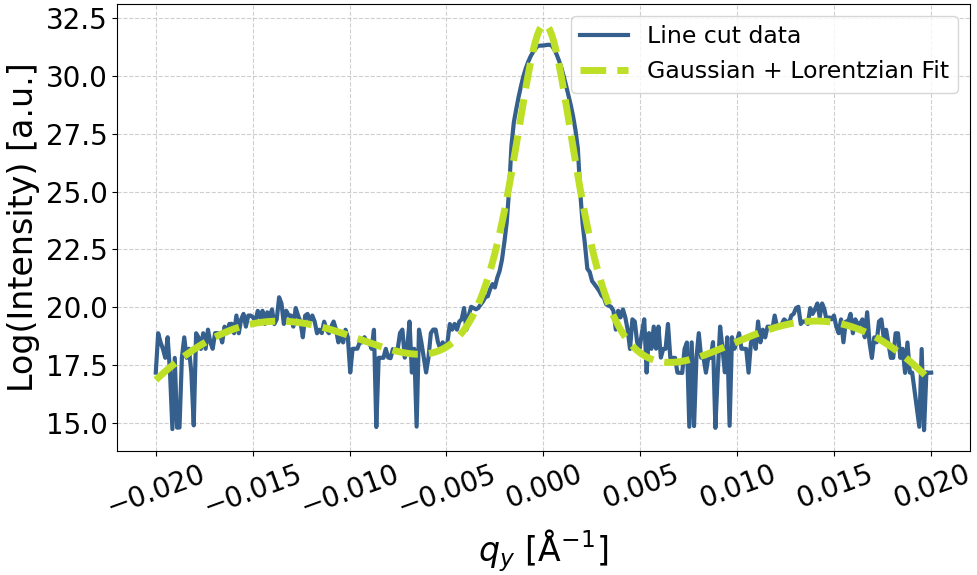}
         \caption{}
         \label{fig:Masksum}
     \end{subfigure}
     
     
     \begin{subfigure}[b]{0.45  \textwidth}
         \centering
         \includegraphics[width=\textwidth]{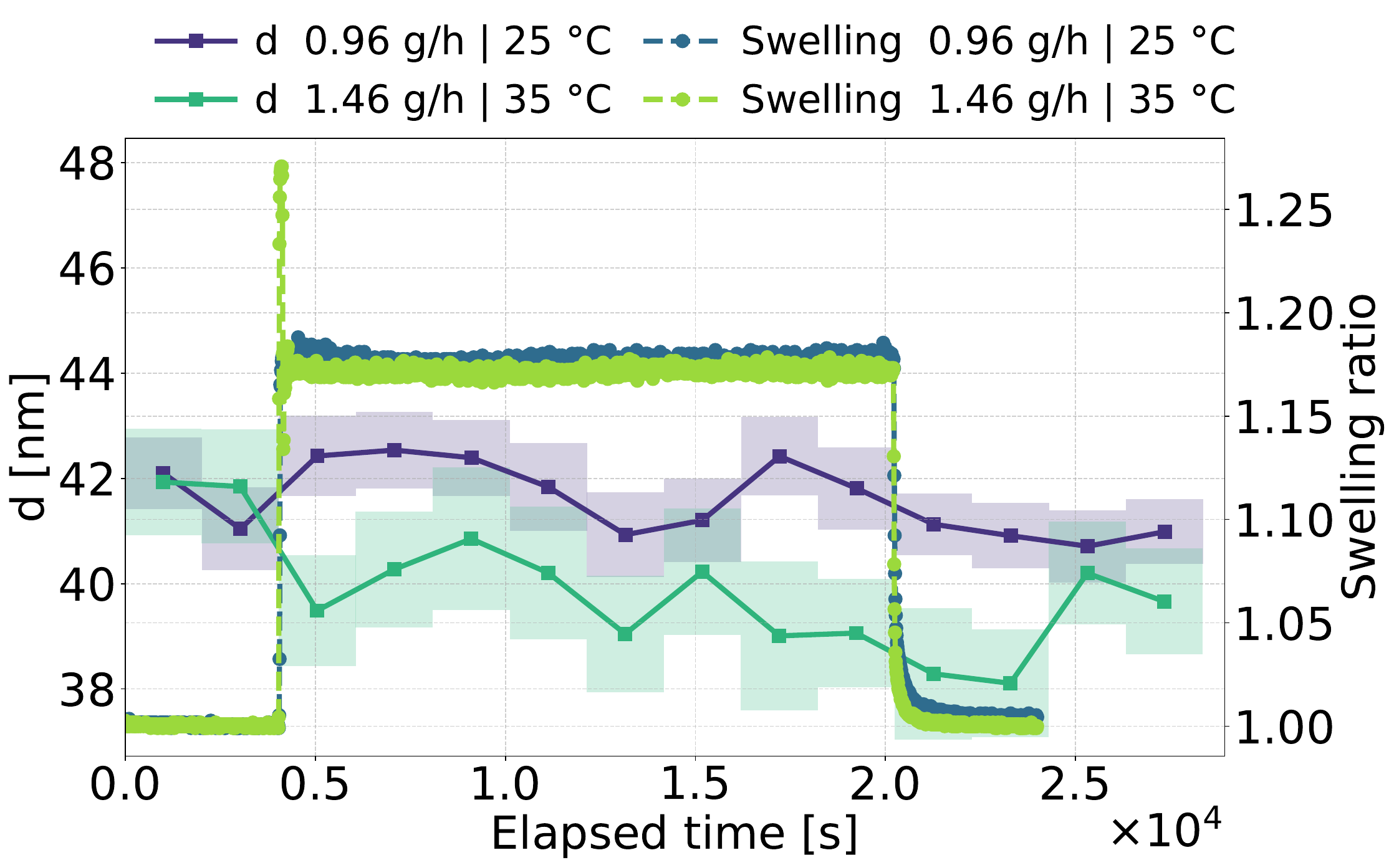}
         \caption{}
         \label{fig:distT}
     \end{subfigure}
     
     \caption{\textit{In situ} GISAXS results during STVA of polystyrene(38)-\textit{b}-poly(ferrocenyl dimethyl silane)(11) thin films. (a) representative GISAXS map during swelling at 25~$^\circ$C with horizontal line-cut box highlighted. Arbitrary units on the color scale bar for the scattered intensity, (b) A representative graph of data from a line-cut fitted to a sum of two Gaussians and a Lorentzian function (Eq.~\ref{GLG}), and (c)  The lateral repeat distance resulting from curve fitting as function of time during annealing (with $\pm$ uncertainty marked as a region around the experimental points) at 25 and 35~$^{\circ}$C. The corresponding swelling ratio is also plotted.}
     \label{fig:Mask}
\end{figure}

This experiment explores BCP structural reorganizing during STVA at two different temperatures and is performed in the magnetic drawer using dummies in place of magnets, i.e., without application of an external magnetic field. Here, the combinations of annealing temperature and solvent gas flow through the chamber were selected using preliminary calibration runs to iteratively adjust the toluene humidity in the chamber until both annealing temperatures produced the same final equilibrium swelling ratio, i.e., a higher temperature requires a higher solvent humidity. Polystyrene(38)-\textit{b}-poly(ferrocenyl dimethyl silane)(11) (PDI 1.25) thin film samples were prepared by spin coating a 3~wt\% BCP solution in toluene onto Si wafers. The BCPs bulk morphology is expected to consist of poly(ferrocenyl dimethyl silane) spheres arranged on a BBC lattice  in a PS matrix \cite{LammertinkThesis}. The two as-spun dry films had thicknesses of 110 and 111~nm and were  designated for annealing at 25 and 35~$^{\circ}$C, respectively. A third control sample with a thickness of 111~nm was prepared under identical conditions and left unannealed. All thicknesses were measured using the spectral reflectometer in the STVA chamber. The two samples were annealed for 4~h and \textit{in situ} GISAXS was used to track the reorganizing of the thin film structure. The equilibrium absorption of toluene vapor was kept the same for the two films annealed at different temperatures by selecting a toluene flow of 0.96 and 1.46~g/h for 25~$^{\circ}$C and 35~$^{\circ}$C, respectively, resulting in a maximal equilibrium swelling ratio of 1.17 in both cases (see ~Fig.~\ref{fig:distT}). The relative toluene humidity in the chamber during annealing was $52\%$ at 25~$^{\circ}$C and $48\%$ at 35~$^{\circ}$C.

\begin{figure}[!h]
    \centering
    \includegraphics[width=0.95\linewidth]{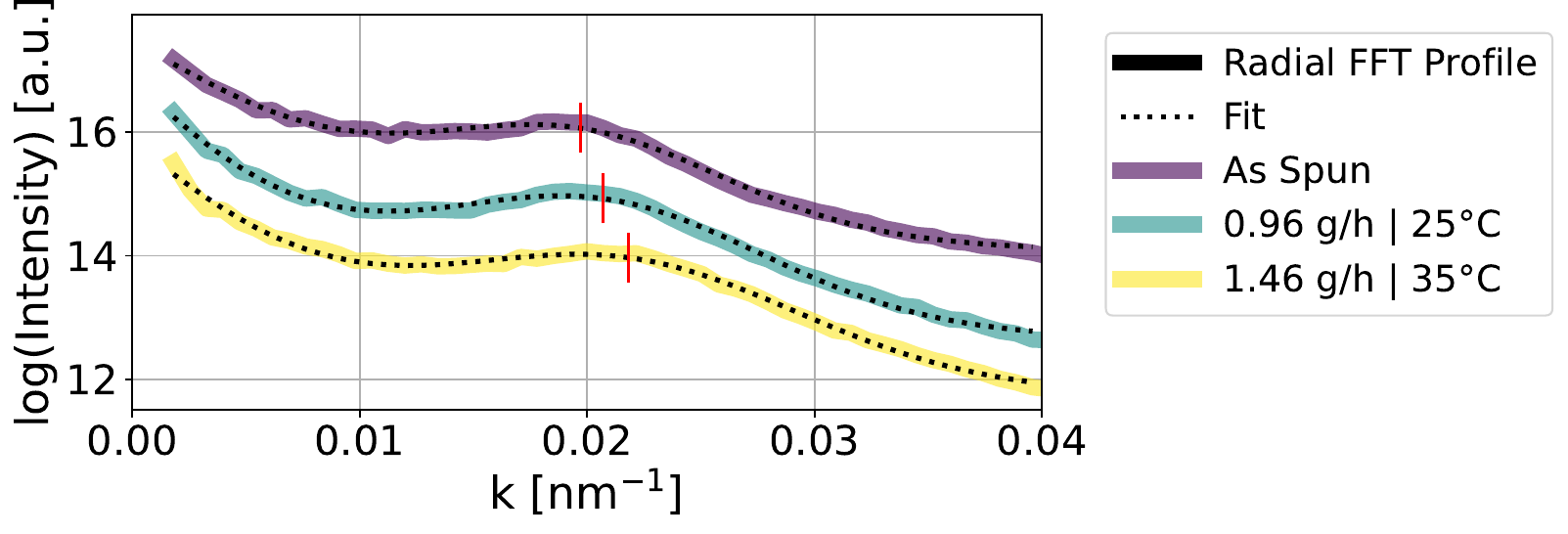}
    \caption{Radial intensity profiles of the Fourier transforms  derived from AFM topography images of PS(38)-\textit{b}-PF2MS(11) films before and after solvent annealing in toluene vapor under different conditions (see legend) with summed Lorentzian + Gaussian fits (offset for clarity). Gaussian peak positions in legend entry order from the top and  marked with a red dash on the curves, are 0.0197, 0.0207, and 0.0218~nm$^{-1}$}
    \label{fig:fft2dradial2}
\end{figure}

AFM topography images acquired before and after STVA were Fourier transformed. The transforms exhibit diffuse rings characteristic of short‐range ordered surface microdomains (See Supplementary Material Fig.~S6). These patterns were then azimuthally averaged to produce radial intensity profiles in reciprocal space, $I(k)$ (Fig.\ref{fig:fft2dradial2}). Each profile was fitted to a Gaussian plus a Lorentzian function to locate the peak positions, $k^*$. Calculating the corresponding repeat distances, $d=1/k^*$, indicates a 50.8~nm repeat distance for the unannealed control sample, while the repeat distances for the samples annealed at 25 and 35~$^{\circ}$C were 48.7 and 45.9~nm, respectively. This reduction in repeat distance is investigated further by examining the \textit{in situ} GISAXS data collected during annealing of these samples.

Using a laboratory X-ray source necessitates longer exposure times than at a synchrotron source. In the present case, the exposure time was 2028~s for a GISAXS map with a negligible interval between successive exposures measured at an incidence angle of 0.18$^\circ$. The high electron-density contrast between the iron-rich minority block and the majority block of the BCP used in this study allows well-developed peaks to be identified in line-cuts, even though features in the corresponding GISAXS maps appear relatively faint. A reduction in the signal-to-noise ratio arises from absorption and parasitic scattering due to the Kapton windows in the beam path. In particular, the Kapton window that separates the sample environment enclosure from the detector vacuum stretches and curves under the one-atmosphere pressure difference, hence adding to the diffuse background. \cite{Antimonov2015}.

The spheres formed by the PS minority block are arranged with a characteristic spacing \textit{d}, giving rise to an intensity maximum located at $q_y=2\pi/d$ in the Yoneda band. The peak is emphasized by summing the intensity in the horisontal box marked on Fig.~\ref{fig:GISAXSmask}. The resulting curve, I($q_y$) (Fig.~\ref{fig:Masksum}), is fitted to a sum of two Gaussian functions corresponding to the lateral peaks and a Lorentzian function to account for the specular contribution: 
\begin{equation} \label{GLG}
\mathrm{log~} I(q_y) = \sum_{i=1}^{2} A_i \exp\left[-\frac{(q_y - q_i)^2}{2\sigma_i^{2}}\right] + \frac{L_0}{1 + \left(\frac{q_y - q_0}{\Gamma}\right)^{2}},
\end{equation}
where $A_i$, $q_i$, and $\sigma_i$ are the amplitudes, center positions, and widths of the two Gaussians, and $L_0$, $q_0$, and $\Gamma$ are the amplitude, center, and half-width at half-maximum of the Lorentzian (Fig.~\ref{fig:Masksum}). The uncertainty in \textit{d} is the standard deviation of the peak positions returned by the fit routine - for the two Gaussians, the average value of \textit{d} and the corresponding uncertainties are used. Fig.~\ref{fig:distT} shows that \textit{d} already in the beginning of the annealing process becomes smaller for the sample annealed at 35~$^{\circ}$C  compared to the sample annealed at 25~$^{\circ}$C (the two samples have the same approximate repeat distance \textit{d} in the dry state before annealing). This is consistent with the observations from the AFM images, which show a slight decrease in the characteristic distance after annealing, most notably for the sample annealed at 35~$^{\circ}$C. However, the repeat distances derived from the AFM images (50.8~nm prior to annealing, 48.7 and 45.9~nm after annealing for 25 and 35~$^{\circ}$C, respectively) are systematically higher than the repeat distances obtained with GISAXS (41.6$\pm$0.7 and 41.9$\pm$1.0~nm prior to annealing and 40.1$\pm$0.7 and 39.9$\pm$1.0~nm after annealing at 25 and 35~$^{\circ}$C, respectively (the GISAXS numbers are found by averaging values from two initial dry GISAXS maps and from two final dry GISAXS maps, respectively)). This difference between AFM and GISAXS results is consistent with the observation in research example B. However, the 'dry' values in the present case are \textit{lower} after annealing than prior to annealing. This result suggest that the final 'dry' microdomain periodicity may depend on the detailed annealing protocol, e.g., with different degrees of strain induced during drying or released during swelling \cite{Sinturel2013}.
 
\FloatBarrier

\subsection{Combined STVA and surface modification using a brush layer}
The final example demonstrates the ability of STVA to induce long-range order in combination with the use of brush layers to modify substrate-polymer interactions. The use of brush layers to facilitate block copolymer self-assembly and improve microdomain ordering has previously been demonstrated \cite{Jung2010, Tavakkoli2012, Dinachali2015}. Here, a brush layer refers to a dense monolayer of short polymer chains covalently grafted, i.e., chemically anchored by one end, to the Si substrate with an oxidized top layer so that the short polymers extend normal to the substrate surface. In our study, three samples were prepared for STVA: two samples with brush layers, one with a PS brush layer and the other with a polydimethylsiloxane (PDMS) brush layer, and one sample without any brush layer. For the two samples designated for brush layer grafting, plasma oxidation (air) of the Si substrate was used to generate surface silanol (Si–OH) groups on the native oxide layer for the $\omega$-hydroxyl end groups of the PS or PDMS chains to covalently anchor to. Immediately afterward, the substrates were spin-coated with a 1~wt\% toluene solution of $\omega$-hydroxyl terminated PS (2.2~kg/mol) or PDMS (5~kg/mol). These samples were subsequently annealed under vacuum  at 150~$^\circ$C for 15~h. Following thermal annealing and brush layer formation, a thin film of PS(28)-\textit{b}-PDMS(85) was spin-coated with a 1.5~wt\% solution in toluene onto each substrate yielding films with a total approximate thickness of 35~nm. The control film was similarly spin-coated onto an untreated Si wafer. Subsequently, the thin films were solvent-annealed at RT for 2~h at 80\% toluene humidity. Cylindrical morphology is expected for this BCP sample\cite{Bates1990}.\\
\begin{figure}[t] 
     \centering
     \begin{subfigure}[b]{0.23\textwidth}
         \centering
         \includegraphics[width=\textwidth]{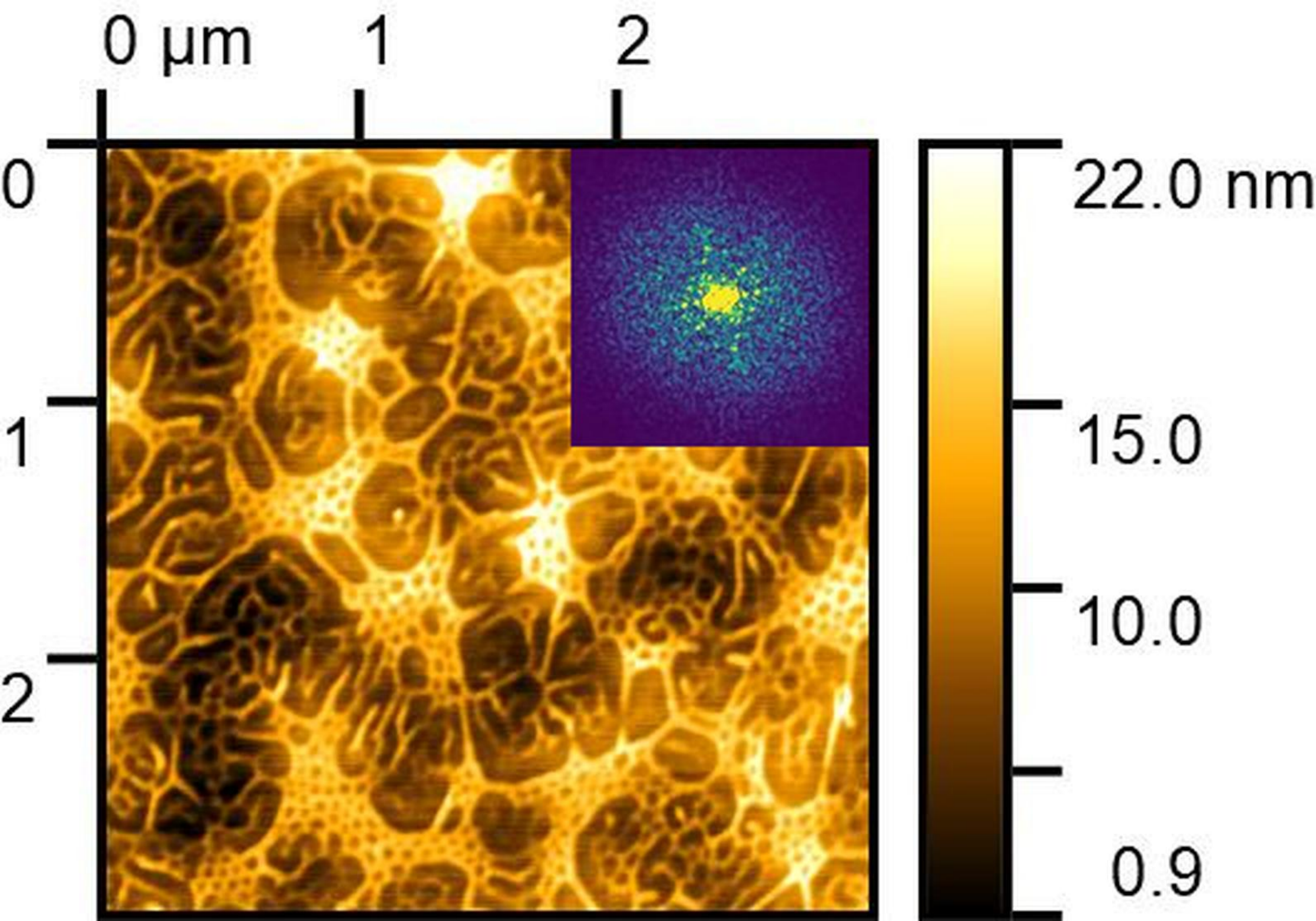}
         \caption{}
         \label{fig:pdmsA}
     \end{subfigure}
     \hfill
     \begin{subfigure}[b]{0.23\textwidth}
         \centering
        \includegraphics[width=\textwidth]{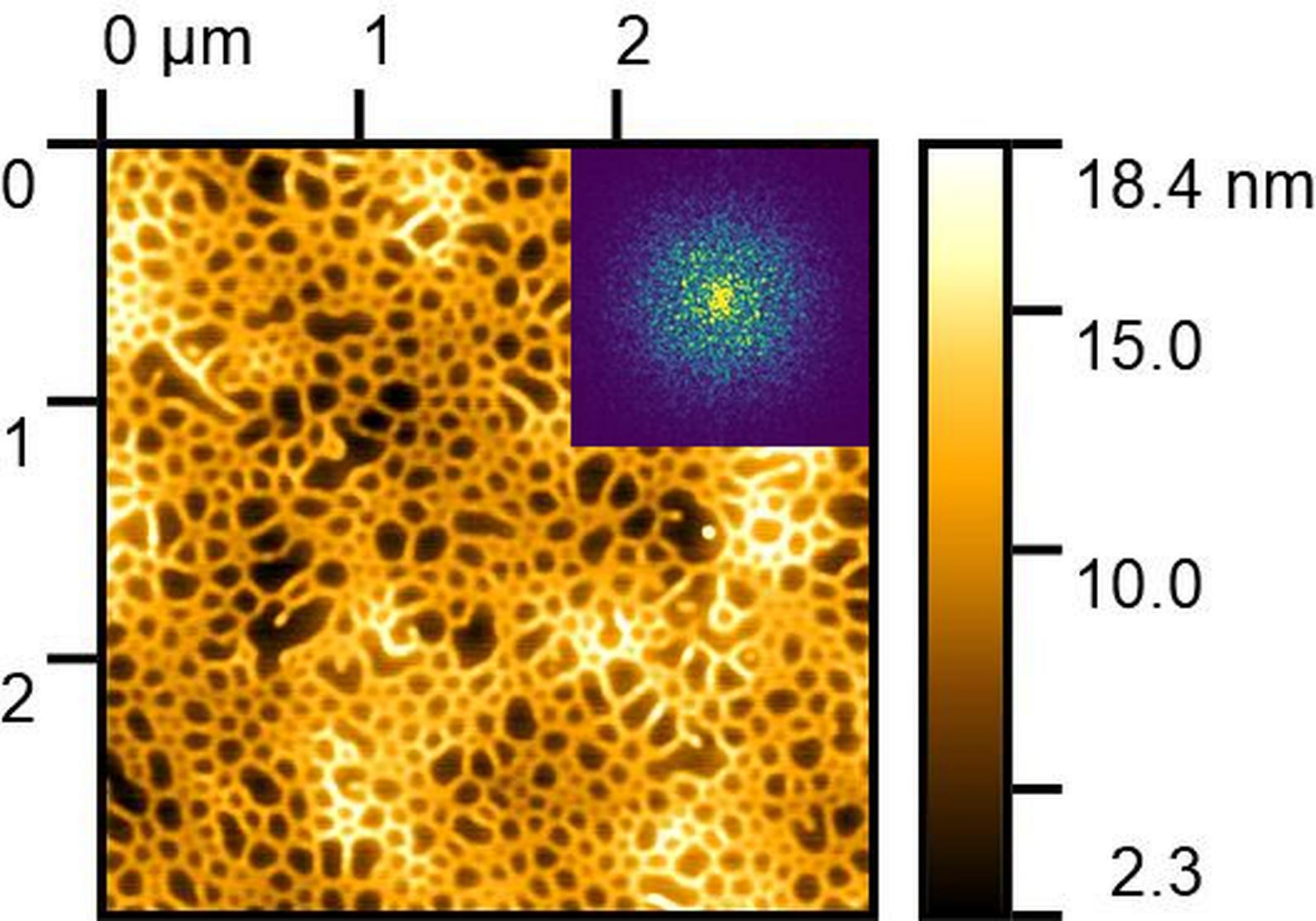}
         \caption{}
         \label{fig:pdmsB}
     \end{subfigure}
     \begin{subfigure}[b]{0.23\textwidth}
         \centering
         \includegraphics[width=\textwidth]{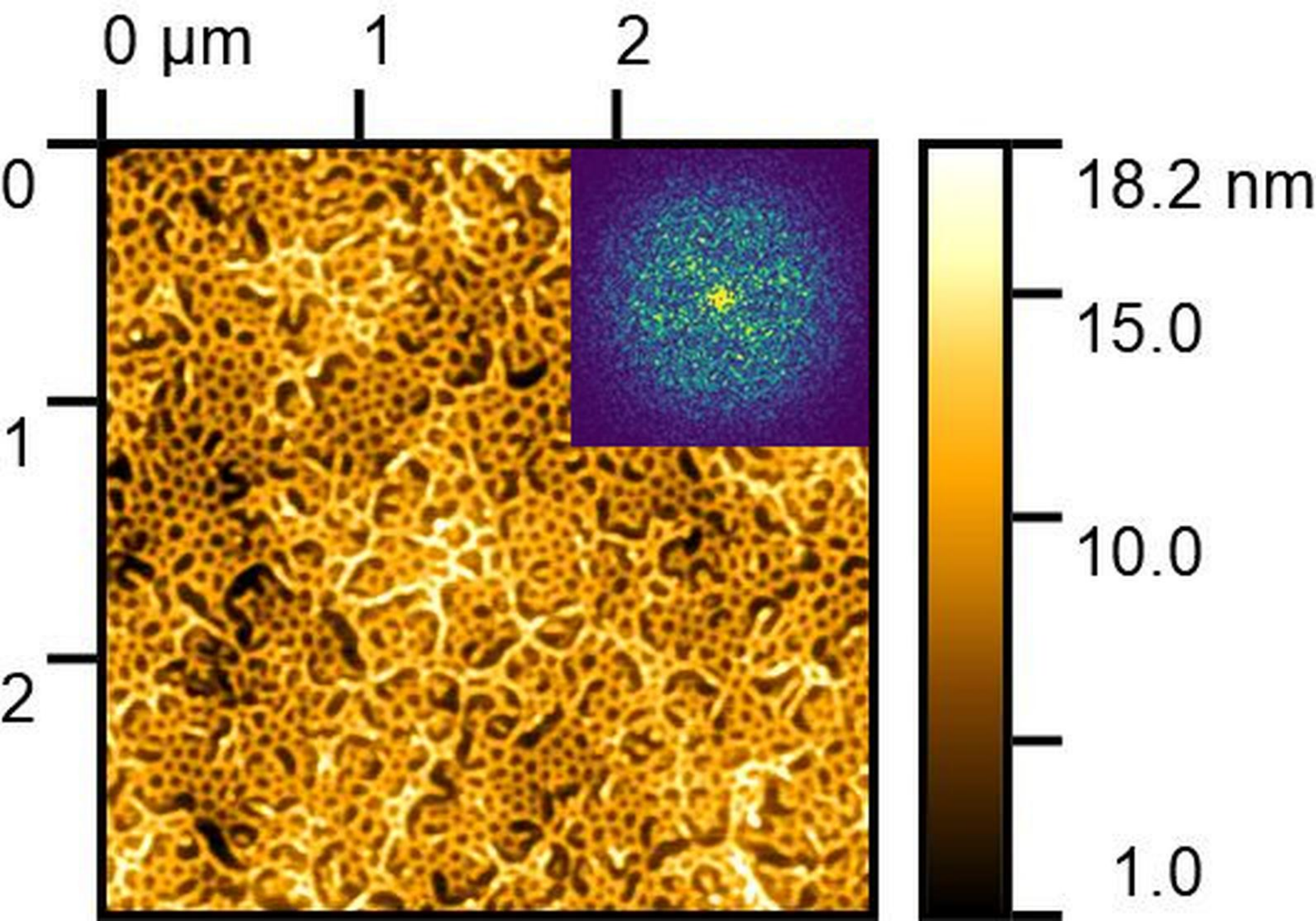}
         \caption{}
         \label{fig:pdmsC}
     \end{subfigure}
     \hfill
     \begin{subfigure}[b]{0.23\textwidth}
         \centering
        \includegraphics[width=\textwidth]{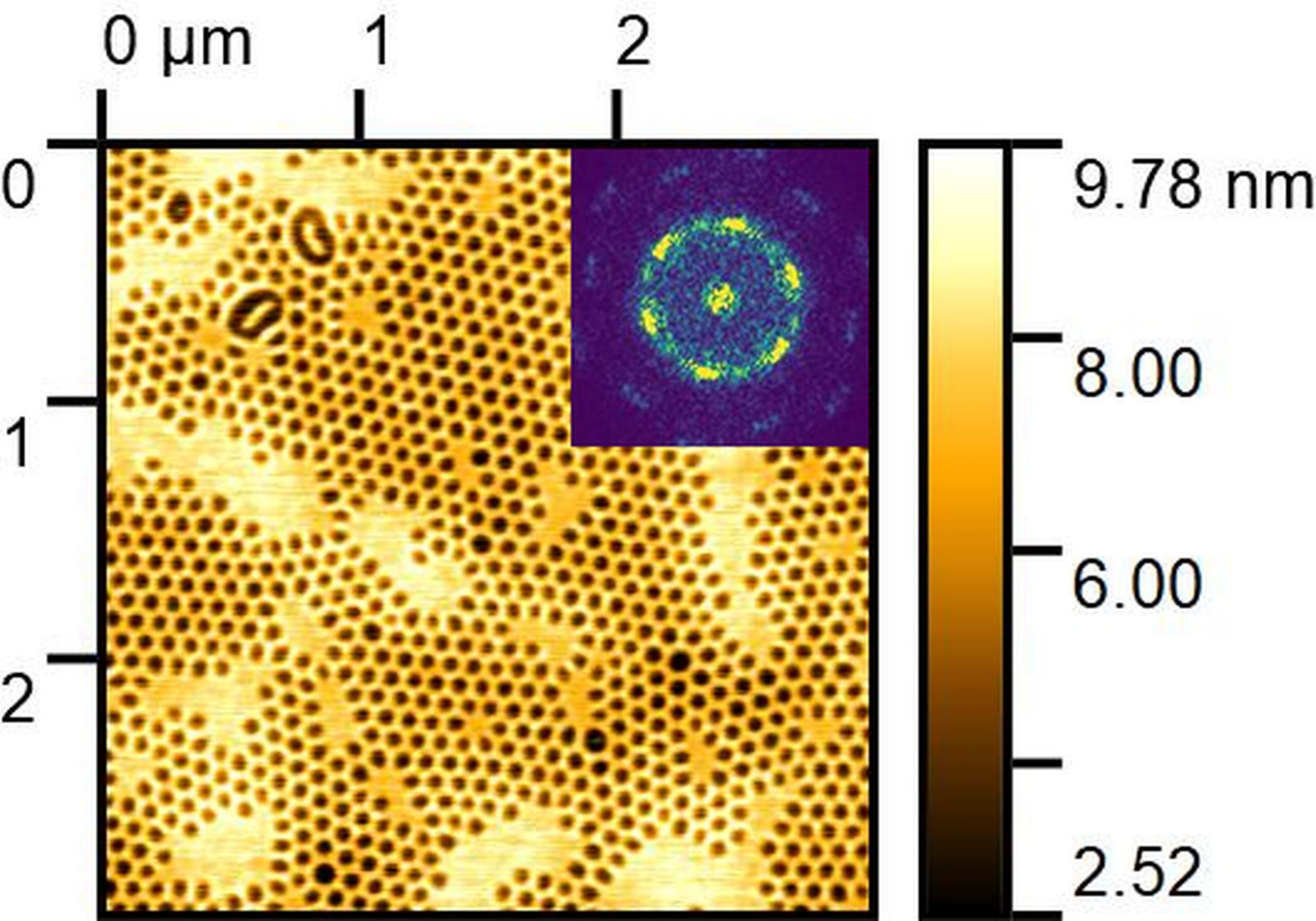}
         \caption{}
         \label{fig:pdmsD}
     \end{subfigure}
     \begin{subfigure}[b]{0.23\textwidth}
         \centering
         \includegraphics[width=\textwidth]{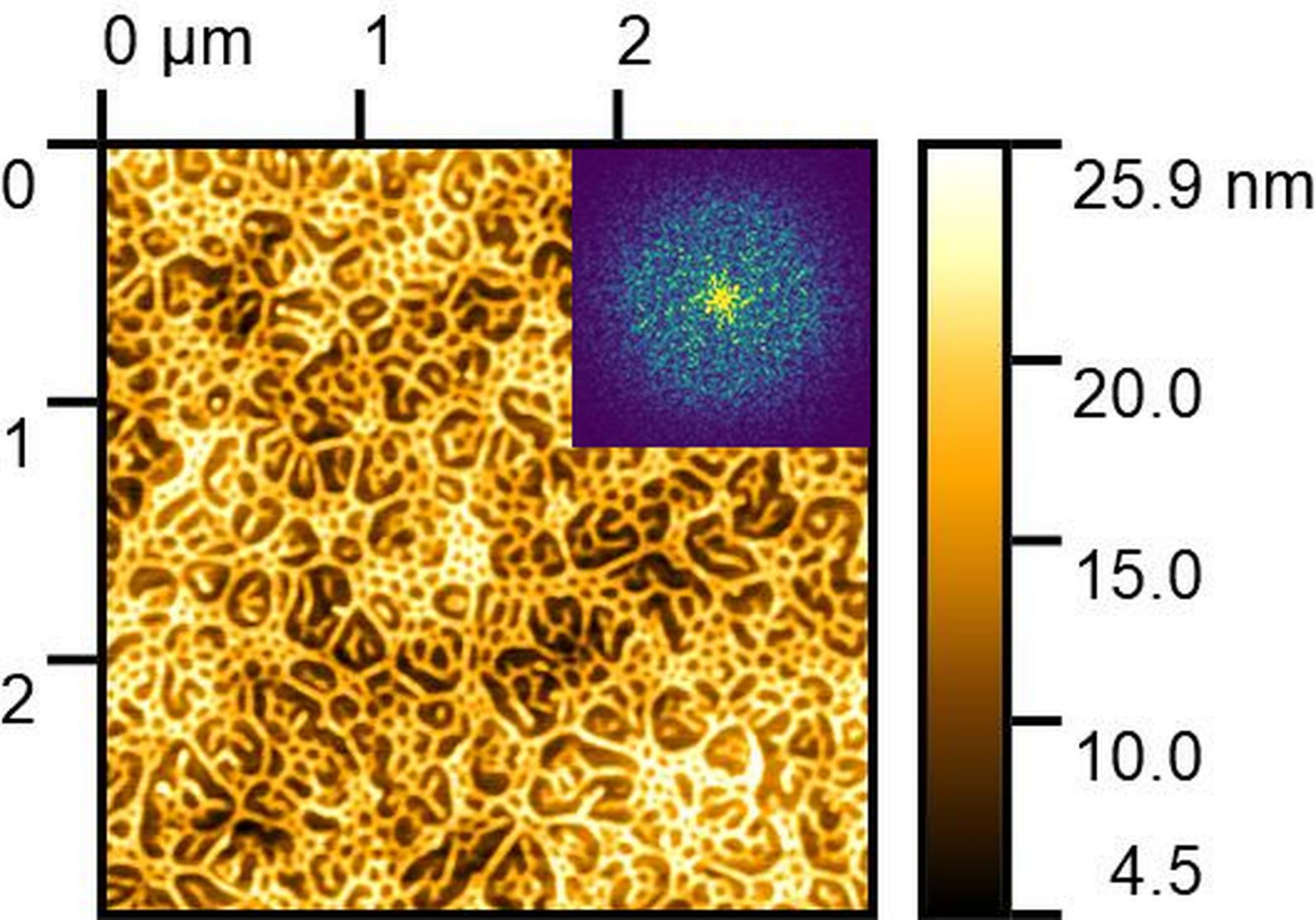}
         \caption{}
         \label{fig:pdmsE}
     \end{subfigure}
     \hfill
     \begin{subfigure}[b]{0.23\textwidth}
         \centering
        \includegraphics[width=\textwidth]{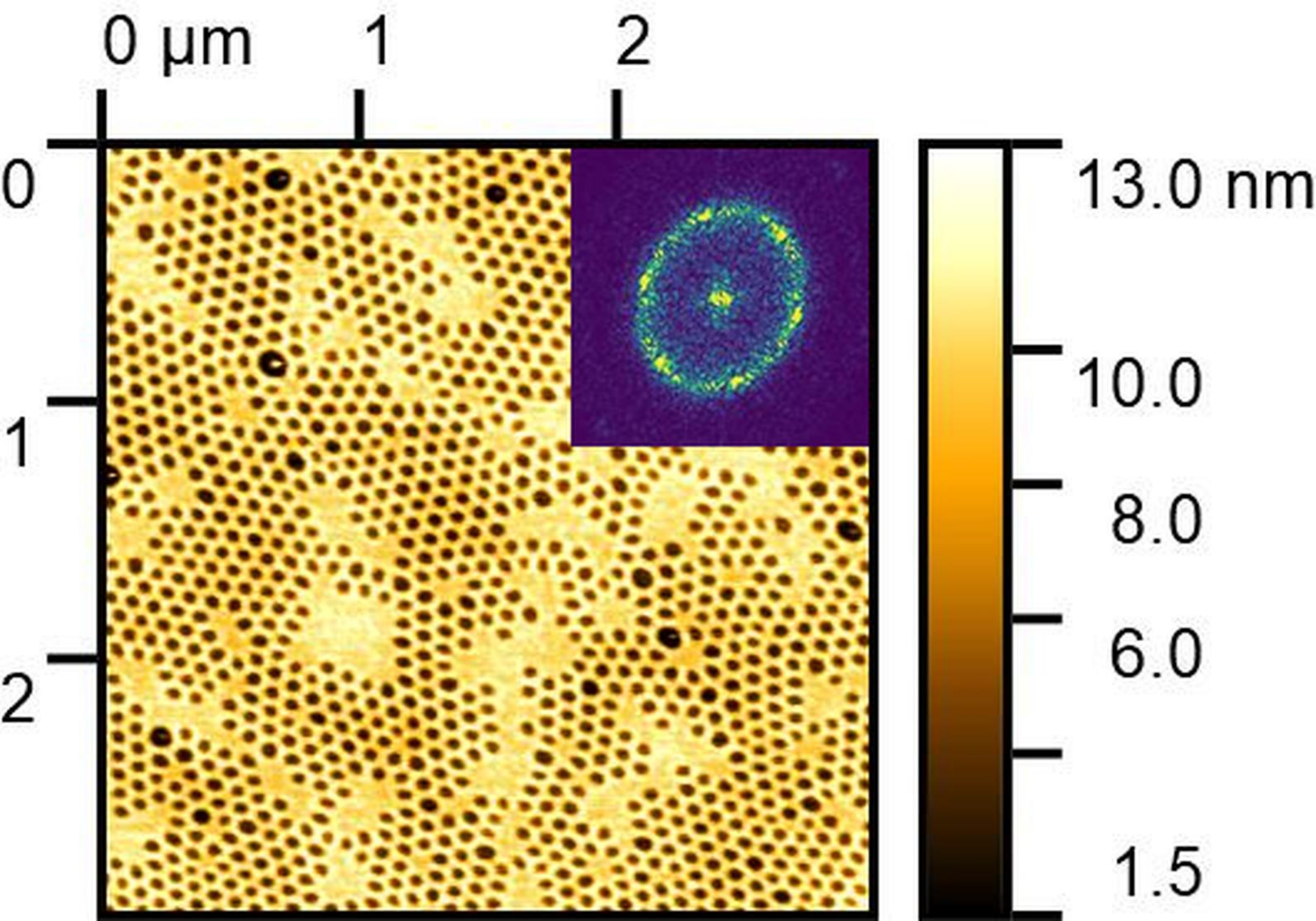}
         \caption{}
         \label{fig:pdmsF}
     \end{subfigure}
        \caption{AFM topography images of 35~nm-thick PS(28)-\textit{b}-PDMS(85) films on Si substrates with and without use of a brush layer. Insets are 2-D Fourier transforms of the corresponding images. (a) control sample, no brush layer, as-spun; (b) control sample, no brush layer, after STVA; (c) with PS brush layer, as-spun; (d) with PS brush layer, after STVA; (e) with PDMS brush layer, as-spun; (f) with PDMS brush layer, after STVA. Hexagonal order, indicated by the six-fold symmetry in the Fourier transform, is observed only in the annealed films using a PS or PDMS brush layer (d,~f).}
        \label{fig:PDMS}
\end{figure}
AFM topography images were used to characterize the microdomain ordering for each sample. The control sample without a brush layer exhibited limited ordering both as-spun (Fig.~\ref{fig:pdmsA}) and after solvent vapor annealing (Fig.~\ref{fig:pdmsB}). Samples with PS and PDMS brush layers showed similarly limited ordering in their as-spun states (Figs.~\ref{fig:pdmsC}~and~\ref{fig:pdmsE}, respectively). After solvent vapor annealing, however, distinct hexagonal arrangements of standing cylinders emerged in both brush-layer samples, as evidenced by the Fourier transform insets (Figs.~\ref{fig:pdmsD}~and~\ref{fig:pdmsF}). These results demonstrate that brush layers substantially enhance ordering in thin BCP films, \textit{in casu} PS-\textit{b}-PDMS thin films, but that solvent annealing is required for this ordering to occur.

\FloatBarrier

\section{Conclusion}
A compact, modular environmental control setup has been developed and validated. It integrates control of solvent vapor humidity in the chamber, temperature regulation, application of adjustable magnetic fields, and is applicable for GISAXS studies - \textit{in situ} and \textit{ex situ}. The modularity of the design facilitates straightforward sample exchange and reconfiguration for instance when swapping standard drawers with magnetic-field drawers. Evaluations of the chamber performance, including fill and quench times and magnetic field distributions by Gauss meter measurements and confirmed by FEM simulations, demonstrate that the setup enables advanced thin film studies. 

The setup is transportable and can be used at both laboratory and synchrotron X-ray sources. Only minor hardware changes are required for other scattering modes: replacing the Kapton windows with thin aluminum foils possibly together with a sample stage accomodating larger samples, can be used to adapt the set‑up for GISANS, whereas simply adding a sample mount to the existing stage suffices for transmission SAXS.

The broad utility of the setup is demonstrated with four examples covering use of the magnetic-field stage to align magnetic nanoparticles in a BCP matrix during annealing; \textit{ex situ} GISAXS of a di-BCP thin film before and after STVA showing the creation of long-range order after annealing; \textit{in situ} GISAXS study of the influence of STVA solvent humidity and annealing temperature on the restructuring of a di-BCP thin film and finally the use of STVA to obtain order in a di-BCP thin film when combining with brush layer surface modification.  Overall, the versatile setup has a large potential for use in engineering of thin film nanoscale structures and hence in e.g., photonics, nanolitography, as sensors and in membrane and filtration technologies.

\begin{acknowledgments}
We thank Jonathan M. Gow, Ib Høst Pedersen, Rolf Skjoldbirk Rønne, Oliver V. Trasbo Madsen, Johnny Faust, Bjarne Christensen, all Roskilde University, for valuable input to the design and use of the STVA setup.
Christian Kjeldbjerg was supported by the Danish Committee for Scientific Infrastructure (NUFI) through the ESS Lighthouse ''Q-MAT: Magnetism and Quantum Materials``. The Novo Nordisk Foundation is gratefully acknowledged for funding RUCSAXS – Roskilde University Interdisciplinary X-ray Scattering Hub – with grant NNF21OC0068491.

The authors have no conflicts of interest to disclose.

\end{acknowledgments}
\bibliography{aipsamp}

\end{document}